\documentclass[twocolumn]{autart}   
\usepackage{amsmath,amssymb,amsfonts}
\usepackage{graphicx}
\usepackage{color}
\usepackage{bbding}
\usepackage{wrapfig}
\usepackage{pifont}
\usepackage{cases}
\usepackage{subfigure}
\usepackage{balance}
\usepackage[T1]{fontenc}
\usepackage{cite}
\usepackage{natbib}
\usepackage[justification=centering]{caption}
\usepackage{graphicx}
\usepackage{algorithm}
\usepackage{algorithmic}
\usepackage{mathtools}

\usepackage{bm}
\usepackage{listings}
\newtheorem{remark}{Remark}

\newtheorem{theorem}{Theorem}
\allowdisplaybreaks
\newtheorem{proposition}{Proposition}
\newtheorem{lemma}{Lemma}
\newtheorem{example}{Example}

\newtheorem{assumption}{Assumption}
\newtheorem{problem}{Problem}
\newcommand{\oomit}[1]{}

\begin{document}

\begin{frontmatter}

\title{Sufficient and Necessary Smooth Barrier-Like Conditions for Continuous-Time Finite-Horizon Safety Verification}

\thanks[footnoteinfo]{
Corresponding author Bai Xue. Tel. +8613552483249.}

\author[]{Bai Xue 
}\ead{xuebai@ios.ac.cn}  

\address{KLSS, Institute of Software, Chinese Academy of Sciences, Beijing, China}   


\begin{abstract}
We study finite-horizon safety verification for continuous-time stochastic systems modeled by stochastic differential equations. Given an open and bounded safe set, a compact initial set, a finite time horizon, and a
prescribed violation-probability threshold, the objective is to certify that the exit probability from the safe set before the horizon remains below the threshold. Under uniform ellipticity of the diffusion matrix and regularity
of the safe-set boundary, we develop a time-dependent barrier-like condition expressed as a set of functional inequalities, obtained by relaxing a parabolic initial-boundary value problem associated with the exit
probability. If a function satisfies the condition, then the probability of leaving the safe set before the horizon is guaranteed to remain below the threshold. Conversely, when the exit probability is strictly smaller than
the prescribed threshold uniformly over the initial set, the barrier-like condition is also necessary: there exists a smooth function satisfying it. Furthermore, when the dynamics are polynomial and the set descriptions are
semialgebraic, the problem of verifying the condition admits a sum-of-squares formulation that is both sound and complete, yielding a complete degree hierarchy for finite-horizon stochastic safety verification.
\end{abstract}

\end{frontmatter}

\maketitle
\section{Introduction}
\label{sec:intro}
Safety is a fundamental requirement in autonomous systems, robotics, and learning-enabled control, where uncertainty and stochastic disturbances may cause system trajectories to leave a prescribed safe region. For continuous-time systems, stochastic differential equations (SDEs) provide a natural framework for modeling the combined effects of deterministic dynamics and stochastic disturbances, typically represented by diffusion terms driven by Brownian motion \cite{oksendal2013stochastic}. Unlike discrete-time systems, for which safety is often evaluated at discrete sampling instants, continuous-time safety verification must account for the entire trajectory over a time interval. This pathwise requirement naturally leads to the study of first exit times and their associated probabilities \cite{bujorianu2007new,bujorianu2009stochastic}. In particular, finite-horizon stochastic safety verification asks whether the probability that a trajectory exits a prescribed safe set before a given time horizon remains below a specified threshold. This problem is particularly relevant to applications in which safety must be maintained throughout a finite operational period, including autonomous navigation, robotic task execution, and safety-critical decision-making.

Barrier-certificate-based methods provide an effective approach to this verification problem \cite{prajna2004safety,prajna2007framework}. Rather than computing the finite-horizon exit probability explicitly, these methods seek an auxiliary function, called a barrier certificate, that satisfies suitable functional inequalities and thereby provides an upper bound on the probability of a safety violation. When the SDE coefficients and the relevant sets admit polynomial descriptions, these inequalities can often be encoded as sum-of-squares (SOS) constraints and solved using semidefinite programming (SDP) \cite{parrilo2003semidefinite}. However, existing barrier-based methods predominantly establish sufficient conditions for safety \cite{santoyo2021barrier,steinhardt2012finite,xue2023new}. Although the existence of a valid certificate guarantees the desired probabilistic safety property, the failure to find such a certificate does not necessarily imply that the specification is violated. In particular, even when the safety requirement is satisfied, the prescribed barrier conditions may admit no feasible certificate. This limitation motivates the development of a converse theory that characterizes not only when a barrier certificate guarantees safety, but also when a suitable certificate necessarily exists.

To address this limitation, we establish sufficient and necessary barrier-like conditions for finite-horizon stochastic safety verification under suitable regularity assumptions on the safe-set boundary and uniform ellipticity of the diffusion matrix. Our approach begins with an analytical characterization of the safety probability. Specifically, we show that the finite-horizon safety probability is characterized by the unique classical solution to an associated parabolic initial-boundary value problem. Building on this PDE characterization, we derive barrier-like conditions by replacing the PDE equality with a system of functional inequalities. We prove that these conditions are sufficient: any suitable function satisfying them guarantees the prescribed upper bound on the finite-horizon exit probability. More importantly, we establish a converse result showing that a suitable polynomial certificate necessarily exists when the exit probability remains strictly below the prescribed threshold uniformly over the initial set. This result connects the probabilistic safety specification directly to the existence of a polynomial satisfying the proposed barrier conditions, thereby overcoming the limitation of purely sufficient barrier-based approaches.

The converse result also provides the foundation for a computationally tractable synthesis framework. When the SDE coefficients are polynomial and the safe and initial sets are semialgebraic, the proposed barrier conditions can be encoded using SOS representations. Under an Archimedean assumption on the associated quadratic modules, Putinar's Positivstellensatz ensures that the required polynomial inequalities admit finite-degree SOS representations under the relevant strict positivity conditions. Consequently, the resulting degree hierarchy is sound and complete under the stated assumptions: feasibility at any degree certifies the desired probabilistic safety guarantee, while every verification instance satisfying the strict safety condition admits a feasible polynomial certificate at some finite degree. Thus, the hierarchy provides more than a sufficient verification procedure: it is guaranteed to succeed whenever the safety specification holds with a strict margin.

The main contributions of this work are threefold:
\begin{enumerate}
\item We introduce a new barrier-like condition for finite-horizon stochastic safety verification and establish its soundness through an analytical characterization of the safety probability using a parabolic PDE.
\item We establish a converse result showing that, under a strict safety margin, a polynomial satisfying the proposed barrier conditions necessarily exists. This result closes the gap between the probabilistic safety specification and the existence of a barrier certificate.
\item We develop an SOS-based synthesis framework for polynomial SDEs and establish soundness and completeness of the associated degree hierarchy under standard algebraic assumptions.
\end{enumerate}

\subsection*{Related Work}
Barrier-certificate-based methods were originally developed for safety
verification of deterministic dynamical systems
\cite{prajna2004safety,prajna2007convex} and were subsequently extended to stochastic systems to provide probabilistic safety guarantees
\cite{prajna2004stochastic,prajna2007framework,steinhardt2012finite,santoyo2021barrier,jagtap2018temporal,liu2019reachability,vzikelic2023learning,abate2024stochastic,laurenti2025unifying,yu2023safe,xue2024sufficient,chen2025construction,xue2021reach,cao2025comparative}. These approaches typically rely on martingale
arguments, including Ville's inequality \cite{ville1939etude}, or on relaxations of equations characterizing the underlying probability \cite{xue2024sufficient,xue2024}. In addition, barrier-certificate-based methods have also been extended to controller synthesis
\cite{wang2021safety,vzikelic2023learning,jagtap2020formal,
kordabad2024control}.

Safety and reach-avoid analysis for continuous-time stochastic systems is closely tied to solving parabolic Kolmogorov equations
\cite{koutsoukos2008computational,bujorianu2007new,bujorianu2009stochastic,esfahani2016stochastic}. Since solving these equations directly can be computationally demanding, barrier-certificate-based-methods adopt a different approach: rather than solving the
PDE, they seek an auxiliary function satisfying suitable barrier-like conditions expressed as functional inequalities. Such inequalities can be addressed by a wide range of computational methods, e.g., \cite{papachristodoulou2005tutorial,peruffo2021automated}. In particular, when the underlying dynamics are polynomial and the relevant sets are semialgebraic,
these inequalities can often be handled efficiently through SOS programming \cite{parrilo2003semidefinite}, which is a convex optimization framework. For bounded-time stochastic safety verification, several barrier-based approaches have been proposed
\cite{steinhardt2012finite,santoyo2021barrier,feng2020unbounded,xue2025refined}. However, these conditions are primarily sufficient: the existence of a barrier certifies safety, whereas the absence of a barrier does not prove that the safety specification is violated. This gap motivates
the development of barrier-like conditions that are both sufficient and necessary.

This distinction motivates a converse characterization. Recent work has established necessary-and-sufficient certificate conditions for discrete-time stochastic systems
\cite{kordabad2026certificates,xue2026sufficient,xue2026converse,xue2026sufficientd} or probabilistic programs \cite{majumdar2025sound}, infinite-horizon reach-avoid verification of continuous-time stochastic systems
\cite{xue2026sufficientc}, and (in)finite-horizon safety verification of continuous deterministic systems
\cite{prajna2007convex,maghenem2022converse,meng2022smooth,li2026converse}. However, a corresponding necessary-and-sufficient characterization for finite-horizon safety verification of systems modeled by SDEs remains underexplored. In this work, we establish sufficient and necessary barrier-like conditions for finite-horizon safety verification under suitable regularity and ellipticity assumptions. Moreover, under a strict safety margin, we show that the barrier can be chosen polynomial, leading to a sound and complete SOS degree hierarchy.

\noindent{\textbf{Notations.}}Throughout this section, for a set $\mathcal{X}$, $\overline{\mathcal X}$ its closure,  $\partial\mathcal X$ its boundary, and $\mathcal{X}^c$ its complement. For sets $K$ and $D$, $K\Subset D$ means that $K$ is compactly contained in $D$, i.e., $K$ is a compact subset of $D$ that stays a positive distance away from the boundary $\partial D$. $B_r(\bm{x}_0)$ denotes the open Euclidean ball of radius $r>0$ centered at $\bm{x}_0$ in $\mathbb{R}^n$, i.e.,
\[
B_r(\bm{x}_0)
=
\{\bm{x}\in\mathbb{R}^n \mid \| \bm{x}-\bm{x}_0\|<r\}.
\] 
Given two nonempty sets \(A,B\subseteq\mathbb R^n\), the Euclidean distance between them is defined as
\[ \operatorname{dist}(A,B) := \inf_{\bm x\in A,\;\bm y\in B} \|\bm x-\bm y\|_2. \]
For function classes, unless the boundary is explicitly included, $D$ denotes an open set and $I$ an open time interval. $C^k(D)$ denotes the space of $k$-times continuously differentiable functions on $D$. For $\alpha\in(0,1)$, $C^{k,\alpha}(D)$, equivalently $C^{k+\alpha}(D)$, denotes the space of functions whose $k$-th derivatives are Hölder continuous with exponent $\alpha$ on $D$. If the regularity is required to hold uniformly up to the boundary, we write $C^{k,\alpha}(\overline D)$ or $C^{k+\alpha}(\overline D)$. For parabolic problems, $C^{2,1}(D\times I)$ denotes the class of functions that are twice continuously differentiable in the spatial variables and once continuously differentiable in time. The space $C^{2,1}_{\mathrm{loc}}(D\times I)$ consists of all functions
$u$ whose spatial derivatives up to order two and time derivative up to
order one exist and are continuous on every compact subset of $D\times I$.
Equivalently, $u\in C^{2,1}_{\mathrm{loc}}(D\times I)
\Longleftrightarrow
u\in C^{2,1}(K)
\quad\text{for every compact }K\Subset D\times I$.
The parabolic Hölder space $C^{2+\alpha,1+\alpha/2}(D\times I)$
consists of functions whose spatial derivatives of order two and
first-order time derivative are Hölder continuous with exponent
$\alpha$ with respect to the parabolic metric
\[
d((x,t),(y,s))=|x-y|+|t-s|^{1/2}.
\]
If the Hölder coefficient is required to be uniform up to the boundary, we write $C^{2+\alpha,1+\alpha/2}(\overline D\times \overline I)$.
In \cite{friedman2008partial}, this boundary-regular space is written $\overline{C}_{2+\alpha}(D\times I)$. Thus $C^{2+\alpha,1+\alpha/2}(D\times I)$ usually denotes interior parabolic Hölder regularity, while $\overline{C}_{2+\alpha}(D\times I)$ indicates uniform regularity on the closure, including the boundary. For a compact subset $K\Subset D\times I$, the parabolic Hölder norm is defined by
\[
\begin{aligned}
\|u\|_{C^{2+\alpha,1+\alpha/2}(K)}
={}&\sum_{|\beta|+2j\le 2}
\sup_{(x,t)\in K}
\left|D_x^\beta\partial_t^j u(x,t)\right|\\
&+\sum_{|\beta|+2j=2}
\left[D_x^\beta\partial_t^j u\right]_{\alpha;K},
\end{aligned}
\]
where
\[
[f]_{\alpha;K}
=
\sup_{\substack{(x,t),(y,s)\in K\&(x,t)\ne(y,s)}}
\frac{|f(x,t)-f(y,s)|}
{\bigl(|x-y|+|t-s|^{1/2}\bigr)^\alpha}.
\]
Here, $D_x^\beta$ denotes a spatial derivative indexed by the
multi-index $\beta$, and $\partial_t^j$ denotes the $j$-th time
derivative. The norm controls the function and its spatial derivatives
of order at most two, its first-order time derivative, and the
parabolic Hölder continuity of its second-order spatial derivatives
and first-order time derivative. For a compact subset $K\Subset D\times I$, the $C^{2,1}$ norm is
defined by
\[
\|u\|_{C^{2,1}(K)}
=
\sum_{|\beta|+2j\le 2}
\sup_{(x,t)\in K}
\left|D_x^\beta \partial_t^j u(x,t)\right|.
\]
The space $C_c^\infty(\mathbb R^{1+n})$ consists of all infinitely
differentiable functions $f:\mathbb R^{1+n}\to\mathbb R$ whose support is
compact:
\[
C_c^\infty(\mathbb R^{1+n})=\left\{
f :  \mathbb R^{1+n}\to\mathbb R \Big|
\begin{split}
&f\in C^\infty(\mathbb R^{1+n})\ \text{and} \\
& \operatorname{supp} f \text{ is compact}
\end{split}
\right\}.
\]
Here the support of $f$ is
\[
\operatorname{supp} f
:=
\overline{\{\bm z\in\mathbb R^{1+n}\mid  f(\bm z)\neq 0\}}.
\]
Equivalently, $f\in C_c^\infty(\mathbb R^{1+n})$ if $f$ is smooth and
vanishes outside some bounded set. $\mathbb{R}[\bm{x}]$ denotes the ring of real-coefficient polynomials in the variables $\bm{x}=(x_1,\dots,x_n)$. We also use $L^\infty$ for essentially bounded measurable functions.
\section{Preliminaries}
In this section, we introduce continuous-time stochastic systems modeled by SDEs and formulate the finite-horizon safety verification problem. We begin by introducing SDEs and the assumptions that ensure the well-posedness of their solutions. We then formulate the finite-horizon safety verification problem and present the assumptions adopted throughout this paper for constructing barrier-like conditions.

We consider the continuous-time stochastic system modeled by SDEs of the following form:
\begin{equation}
\label{SDE}
d\bm{X}(t)=\bm{b}(\bm{X}(t))dt +
\bm{\sigma}(\bm{X}(t))d\bm{W}(t),
\end{equation}
where $\bm{W}=(W_1,\ldots,W_k)^\top$ is a standard $k$-dimensional
Wiener process defined on a probability space
$(\Omega,\mathcal{F},\mathbb{P})$. The expectation with respect to $\mathbb{P}$ is denoted by $\mathbb{E}[\cdot]$.

We impose the following standard assumptions on the drift and diffusion coefficients.

\begin{assumption}[Local well-posedness]
\label{assump:sde_conditions}
The drift $\bm{b}:\mathbb R^n\to\mathbb R^n$ and diffusion
$\bm{\sigma}:\mathbb R^n\to\mathbb R^{n\times k}$ are locally Lipschitz continuous; that is, for every compact set $K\subset\mathbb R^n$, there
exists $L_K>0$ such that
\[
\|\bm{b}(\bm{x})-\bm{b}(\bm{y})\|
+
\|\bm{\sigma}(\bm{x})-\bm{\sigma}(\bm{y})\|_F
\leq
L_K\|\bm{x}-\bm{y}\|, \forall\bm{x},\bm{y}\in K,
\]
where $\|\cdot\|_F$ denotes the Frobenius norm.
\end{assumption}

Under Assumption \ref{assump:sde_conditions}, for every initial state $\bm{x}_0\in\mathbb R^n$, the SDE \eqref{SDE} admits a unique maximal strong solution, denoted by $\bm{X}_{\bm{x}_0}(t)$ with $\bm{X}_{\bm{x}_0}(0)=\bm{x}_0$. In the analysis below, we only consider this solution up to the first exit time from the bounded continuation domain introduced below. 

For a function $v:\mathbb{R}^n \times [0,\infty)\to\mathbb{R}$ that is
once continuously differentiable in time and twice continuously
differentiable in space, the infinitesimal generator associated with
\eqref{SDE} is defined by
\begin{equation}
\label{infi}
\mathcal{L}v(\bm{x},t)
=
\lim_{h\downarrow 0}
\frac{
\mathbb{E}\!\left[
v(\bm{X}_{\bm{x}}(h),t+h)
\right]
-v(\bm{x},t)
}{h},
\end{equation}
whenever the limit exists.

\begin{proposition}
\label{prop:inf_generator}
Consider the system~\eqref{SDE} under
Assumption~\ref{assump:sde_conditions}. For every
$v\in C^{2,1}(\mathbb{R}^n\times \mathbb{R})$, the infinitesimal
generator is given by $\mathcal{L}v(\bm{x},t)
=\partial_t v(\bm{x},t)
+\nabla_{\bm{x}}v(\bm{x},t)^\top\bm{b}(\bm{x})+\frac{1}{2}\operatorname{tr}\left(\bm{\sigma}(\bm{x})^\top \bm{H}_{\bm{x}}v(\bm{x},t)\bm{\sigma}(\bm{x})
\right)$, where $\bm{H}_{\bm{x}}v(\bm{x},t)
:=\nabla_{\bm{x}}^2v(\bm{x},t)$ denotes the Hessian matrix of
$v$ with respect to $\bm{x}$. Equivalently, defining
\[
\bm{a}(\bm{x})
:=\bm{\sigma}(\bm{x})\bm{\sigma}(\bm{x})^\top,
\]
the generator can be expressed as
\[
\begin{split}
\mathcal{L}v(\bm{x},t)
=\partial_t v(\bm{x},t)
&+\nabla_{\bm{x}}v(\bm{x},t)^\top\bm{b}(\bm{x})\\
&+\frac{1}{2}\operatorname{tr}\left(
\bm{a}(\bm{x})\bm{H}_{\bm{x}}v(\bm{x},t)
\right).
\end{split}
\]
Moreover, whenever $v$ satisfies the regularity and integrability
conditions required for Dynkin's formula, then for every almost surely finite stopping time $\tau$,
\begin{equation*}
\label{dynkin}
\mathbb{E}
\left[
v(\bm{X}_{\bm{x}_0}(\tau),\tau)
\right]
=
v(\bm{x}_0,0)+ \mathbb{E}
\left[
\int_0^\tau
\mathcal{L}v(\bm{X}_{\bm{x}_0}(s),s)ds
\right].
\end{equation*}
\end{proposition}

See, e.g.,~\cite{oksendal2013stochastic}.

In this paper, we consider a finite-horizon safety verification problem over the time horizon $[0,T]$, where $T<\infty$. Let $\mathcal{X}\subset\mathbb{R}^n$ be a safe set, and
$\mathcal{X}_0\subset\mathcal{X}$ a set of initial states. For $\bm{x}_0\in\mathcal{X}_0$,
define the safe-set exit time by
\[
\tau_{\mathcal{X}}(\bm{x}_0):=
\inf\left\{
t\geq 0\mid
\bm{X}_{\bm{x}_0}(t)\notin\mathcal{X}
\right\}.
\]
The corresponding finite-horizon exit probability is
\begin{equation}
\label{eq:ra_probability}
\mathbb{P}_{\mathrm{exit}}(\bm{x}_0):=
\mathbb{P}
\left(
\tau_{\mathcal{X}}(\bm{x}_0)\leq T
\right).
\end{equation}

\begin{problem}[Finite-horizon Safety Verification]
\label{un_proII}
Given a prescribed probability threshold $\epsilon \in (0,1)$, the finite-horizon safety verification problem is to establish 
\[
\mathbb{P}_{\mathrm{exit}}(\bm{x}_0)\leq\epsilon,
\qquad
\forall\bm{x}_0\in\mathcal{X}_0.
\]
\end{problem}

We study a necessary and sufficient barrier-like condition of Problem \ref{un_proII} in terms of polynomial barrier functions. The following assumptions specify the system and set-theoretic conditions used throughout the paper.

\begin{assumption}
\label{assm2}
\begin{enumerate}
\item The initial set $\mathcal{X}_0$ is compact. The safe set $\mathcal{X}$ is bounded and open, and $\partial\mathcal{X}$ is a compact $C^{2+\gamma}$ hypersurface.

    \item The diffusion covariance matrix  $\bm{a}$ is uniformly elliptic on $\overline{\mathcal{X}}$: there exists $\lambda>0$
    such that
    \[
    \bm{\xi}^\top
    \bm{a}(\bm{x})
    \bm{\xi}
    \geq
    \lambda\|\bm{\xi}\|^2,
    \qquad
    \forall\bm{x}\in\overline{\mathcal{X}},
    \quad
    \forall\bm{\xi}\in\mathbb{R}^n.
    \]
   This is a uniform non-degeneracy condition on the diffusion coefficient.
\end{enumerate}
\end{assumption}
\section{Barrier-like Conditions}
\label{sec:barrier}

In this section we develop sufficient and necessary barrier-like conditions for finite-horizon safety verification. We first establish the classical solvability and regularity of the finite-time safety probability, which is linked to a parabolic initial-boundary value problem. We then introduce a polynomial approximation result in the $C^{2,1}$ topology. Finally, we
establish the main result: a sufficient and necessary barrier-like condition obtained by relaxing the parabolic problem. The existence of a barrier function is sufficient for finite-horizon safety verification, and under a strict safety margin such a barrier function is also necessary. The converse construction
further shows that the barrier function can be chosen to be polynomial.

\subsection{Classical Solvability of Safety Probabilities}
\label{sec:classical-exit}
This subsection establishes the classical solvability and regularity properties of finite-horizon safety probabilities, which are essential for the subsequent converse construction. We first derive a uniform short-time exit-probability bound to characterize the behavior of the safety probability near the initial time. We then establish its classical solvability and positive-time parabolic Hölder regularity up to the spatial boundary, despite the incompatibility between the initial and lateral boundary conditions at $t=0$.

\begin{lemma}[Short-time exit probability]
\label{lem:short-time-exit}
For $\rho>0$, set $K:=\bigl\{\bm{x}\in\mathcal X \mid \operatorname{dist}(\bm{x},\partial\mathcal X)\ge\rho\bigr\}$. Then there exists $\delta_0>0$ such that for all $0<\delta\le\delta_0$,
\begin{equation}
\label{eq:short-time-bound}
\mathbb{P}(\tau_{\mathcal X}\le\delta)
\le
2n\exp\!\left(-\frac{\rho^2}{8nS^2\delta}\right),
\qquad
\forall\bm{x}\in K,
\end{equation}
where $S:=\sup_{\bm z\in\overline{\mathcal X}}\|\sigma(\bm z)\|_F<\infty$. Consequently,
\[
\lim_{\delta\downarrow0}\sup_{\bm{x}\in K}\mathbb{P}(\tau_{\mathcal X}\le\delta)=0.
\]
\end{lemma}
\begin{pf}
The proof is shown in Appendix A. \qed
\end{pf}

The preceding lemma establishes that the exit probability over a short time interval converges uniformly to zero for initial states in compact subsets of $\mathcal X$. Equivalently, the safety probability converges locally uniformly to its initial value $1$. This estimate supports the following regularity result, which characterizes the safety probability as a classical solution of the parabolic initial-boundary value problem and establishes its parabolic Hölder regularity up to the spatial boundary on every time interval bounded away from $t=0$. Despite the incompatibility between the interior initial condition and the lateral boundary condition at $t=0$, these regularity properties provide a key analytical ingredient for the converse construction.

\begin{lemma}
\label{lemma:boundary}
Let $\tau_{\mathcal{X}}(\bm{x}_0) \equiv \inf \{ t \ge 0 \mid \bm{X}_{\bm{x}_0}(t) \notin \mathcal{X} \}$ be the first exit time, and define the exit probability before the time instant $t$ by:
\[
\mathbb{P}_{\mathrm{exit}}(\bm{x}, t) := \mathbb{P}(\tau_{\mathcal{X}}(\bm{x}) \le t), \quad \forall (\bm{x}, t) \in \overline{\mathcal{X}} \times [0, \hat{T}],
\]
where $\hat{T}>0$ is an arbitrary but fixed scalar value.

Under Assumptions \ref{assump:sde_conditions} and \ref{assm2}, the following conclusions hold: 
\begin{enumerate}
    \item[(i)] The safety probability $u(\bm{x},t) = 1-\mathbb{P}_{\mathrm{exit}}(\bm{x}, t)$ is the unique bounded classical solution in 
\[
u\in C^{2,1}\bigl(\mathcal{X}\times(0,\hat T]\bigr)
\cap
C\bigl(\overline{\mathcal{X}}\times(0,\hat T]\bigr)\cap C\bigl(\mathcal{X}\times[0,\hat T]\bigr),
\]
to the parabolic initial-boundary value problem:
\begin{equation}
\label{eq:survival_pde}
\begin{cases}
\partial_t u(\bm{x},t)=Lu(\bm{x},t),
&(\bm{x},t)\in\mathcal{X}\times(0,\hat T],\\
u(\bm{x},t)=0,
&(\bm{x},t)\in\partial\mathcal{X}\times(0,\hat T],\\
\displaystyle\lim_{t\downarrow0}u(\bm{x},t)=1,
&\bm{x}\in\mathcal{X},\\
u(\bm{x},0)=1, & \bm{x}\in \mathcal{X}.
\end{cases}
\end{equation}
\item[(ii)] For every $\delta\in(0,\hat T)$,
\[
u\in
C^{2+\alpha,\,1+\alpha/2}
\bigl(\overline{\mathcal{X}}\times[\delta,\hat T]\bigr)
\]
where $\alpha\in (0,\gamma]$ and $L$ denotes the spatial infinitesimal generator
\[
Lv(\bm{x},t)
=
\nabla_{\bm{x}}v(\bm{x},t)^\top\bm b(\bm{x})
+
\frac12
\operatorname{tr}\!\bigl(\bm a(\bm{x})\bm H_{\bm{x}}v(\bm{x},t)\bigr).
\]
\end{enumerate}
\end{lemma}
\begin{pf}
The proof is shown in Appendix A. \qed
\end{pf}

\subsection{Polynomial Approximation in $C^{2,1}$}
\label{sec:poly-approx}
The sufficient and necessary barrier-like condition in this work involves a barrier function together with its first time derivative, spatial gradient, and spatial Hessian. In the necessity direction, however, the barrier function constructed from the solution in Lemma~\ref{lemma:boundary} is initially only in $C^{2,1}$, not polynomial. To obtain a polynomial certificate, we require that such a function can be approximated, together with its derivatives, by polynomials on compact subsets. The following density lemma provides precisely this tool.

\begin{lemma}[Polynomial approximation in $C^{2,1}$]
\label{lem:poly-approx}
Let $U\subset\mathbb R^{1+n}$ be open and let $K\subset U$ be compact. Write a point of $\mathbb R^{1+n}$ as $(\bm{x},t)$. Suppose $f\in C^{2,1}(U)$. Then there exists a sequence of polynomials $p_k\in\mathbb R[\bm{x},t]$ such that
\begin{equation}
\label{eq:c12-poly-approx}
\|p_k-f\|_{C^{2,1}(K)}
\longrightarrow 0,
\end{equation}
where
\begin{equation}
\label{eq:c12-norm}
\|g\|_{C^{2,1}(K)}:=
\max\left\{
\begin{split}
&\sup_{(\bm{x},t)\in K}|g(\bm{x},t)|,\\
&\sup_{(\bm{x},t)\in K}|\partial_t g(\bm{x},t)|,\\
&\sup_{(\bm{x},t)\in K}\|\nabla_{\bm{x}}g(\bm{x},t)\|,\\
&\sup_{(\bm{x},t)\in K}\|\bm H_{\bm{x}}g(\bm{x},t)\|_F
\end{split}
\right\}.
\end{equation}
\end{lemma}
\begin{pf}
The proof is shown in Appendix A. \qed
\end{pf}



\subsection{Sufficient and Necessary Barrier-like Conditions}
\label{sec:main-barrier}

We now state the main result of this paper: sufficient and necessary (S.N.) barrier-like condition.

Before presenting the main result, we establish a short-time exit-probability estimate that will be used to prove the sufficiency of the barrier-like condition.

\begin{theorem}[S.N. Barrier-like Conditions]
\label{thm:barrier}
Assume Assumptions~\ref{assump:sde_conditions} and~\ref{assm2} hold.
Fix a horizon $T>0$ and a compact set $\mathcal{X}_0\subset\mathcal{X}$
of initial states.

\noindent\textbf{(i) Sufficiency.}
If there exists a function $v\in C^{2,1}\bigl(\overline{\mathcal{X}}\times [0,T]\bigr)$ satisfying
\begin{equation}
\label{eq:barrier}
\begin{cases}
\dfrac{\partial v}{\partial t}-Lv\geq 0,
& (\bm{x},t)\in\overline{\mathcal{X}}\times [0,T],\\[6pt]
v(\bm{x},t)\geq 1,
& (\bm{x},t)\in\partial\mathcal{X}\times[0,T],\\[4pt]
v(\bm{x},0)\geq 0,
& \bm{x}\in\overline{\mathcal{X}},\\[4pt]
v(\bm{x},T)\leq \epsilon,
& \bm{x}\in\mathcal{X}_0,
\end{cases}
\end{equation}
then $\mathbb{P}_{\mathrm{exit}}(\bm{x}_0)\le\epsilon, 
\forall\bm{x}_0\in\mathcal{X}_0$.

\noindent\textbf{(ii) Necessity.}
Conversely, if
\begin{equation}
\label{eq:strict-safety}
\mathbb{P}_{\mathrm{exit}}(\bm{x}_0)<\epsilon,
\qquad
\forall\bm{x}_0\in\mathcal{X}_0,
\end{equation}
then there exists a polynomial $v\in\mathbb{R}[\bm{x},t]$ satisfying
the strict barrier-like condition~\eqref{eq:barrier}, meaning that all four inequalities in~\eqref{eq:barrier} hold strictly.
\end{theorem}
\begin{pf}
\textbf{Notation.}
Let
\[
p(\bm{x},t)=\mathbb{P}_{\mathrm{exit}}(\bm{x}, t),
\qquad
(\bm{x},t)\in \overline{\mathcal{X}} \times [0,T],
\]
denote the exit probability before the time instant $t$. By
Lemma \ref{lemma:boundary}, $p$ is the unique classical solution of
\begin{equation}
\label{eq:p-PDE}
\begin{cases}
\partial_t p-Lp=0, & (\bm{x},t)\in \mathcal{X}\times(0,T],\\
p(\bm{x},t)=1, & (\bm{x},t)\in\partial\mathcal{X}\times(0,T],\\
\lim_{t\downarrow 0}p(\bm{x},t)=0, & \bm{x}\in\mathcal{X},\\
p(\bm{x},0)=0, & \bm{x}\in\mathcal{X},
\end{cases}
\end{equation}
and, for every $\delta>0$ with $\delta<T$,
\begin{equation}
\label{eq:p-regularity}
p\in C^{2+\alpha,1+\alpha/2}
\bigl(\overline{\mathcal{X}}\times[\delta,T]\bigr).
\end{equation}
The parabolic boundary of the cylinder $\mathcal{X}\times(0,T)$ is
\[
\partial_p\bigl(\mathcal{X}\times(0,T)\bigr)
=
\bigl(\partial\mathcal{X} \times (0,T)\bigr)
\;\cup\;
\bigl(\overline{\mathcal{X}} \times \{0\}\bigr).
\]

\noindent\textbf{Part (i): Sufficiency.}

We prove $p\le v$ on $\overline{\mathcal{X}}\times(0,T]$.

\emph{Step 1: Perturbation and boundary layer.}
Fix $\eta>0$ and set
\[
w_\eta(\bm{x},t):=v(\bm{x},t)-p(\bm{x},t)+\eta.
\]
By uniform continuity of $v$ on the compact set
$\overline{\mathcal{X}}\times[0,T]$ and the boundary condition
$v\ge 1$ on $\partial\mathcal{X}\times[0,T]$, there exists
$\rho>0$ such that
\begin{equation}
\label{eq:collar}
v(\bm{x},t)\ge 1-\eta \text{~whenever }
\operatorname{dist}(\bm{x},\partial\mathcal{X})<\rho
\end{equation}
for any $t\in[0,T]$.

\smallskip
\emph{Step 2: Bottom-face control.}
Fix $\delta\in(0,T)$ and consider the truncated cylinder
$\mathcal{X}\times(\delta,T]$. Define
\[
\begin{split}
&K:=\{\bm{x}\in\mathcal{X} \mid \operatorname{dist}(\bm{x},\partial\mathcal{X})\ge\rho\},\\
&C:=\{\bm{x}\in\mathcal{X} \mid \operatorname{dist}(\bm{x},\partial\mathcal{X})<\rho\}.
\end{split}
\]

\emph{On $K$:} By Lemma \ref{lem:short-time-exit}, $p(\cdot,\delta)\to 0$ uniformly on $K$ as
$\delta\downarrow0$. Since $v$ is continuous on
$\overline{\mathcal{X}}\times[0,T]$, $v(\bm{x},\delta)\to v(\bm{x},0)\ge 0$
uniformly on $K$. Therefore, for $\delta$ sufficiently small,
\[
w_\eta(\bm{x},\delta)=
v(\bm{x},\delta)-p(\bm{x},\delta)+\eta\ge 0
\qquad\text{on }K.
\]

\emph{On $C$:} By \eqref{eq:collar},
$v(\bm{x},\delta)\ge 1-\eta$, and since $p(\bm{x},\delta)\le 1$,
\[
w_\eta(\bm{x},\delta)=
v(\bm{x},\delta)-p(\bm{x},\delta)+\eta\ge
(1-\eta)-1+\eta=0
\quad\text{on }C.
\]

Combining the two regions, we have
\begin{equation}
\label{eq:bottom-face}
w_\eta\ge 0
\qquad\text{on }\mathcal{X}\times\{\delta\}.
\end{equation}

\emph{Step 3: Weak minimum principle.}
The parabolic boundary of the truncated cylinder
$\mathcal{X}\times(\delta,T]$ is
\[
\partial_p\bigl(\mathcal{X}\times(\delta,T]\bigr)=
\bigl(\partial\mathcal{X}\times(\delta,T]\bigr)
\;\cup\;
\bigl(\mathcal{X}\times\{\delta\}\bigr).
\]
On the lateral part $\partial\mathcal{X}\times(\delta,T]$, $p=1$ and
$v\ge 1$, so
\[
w_\eta
=
v-p+\eta
\ge
1-1+\eta
=
\eta>0.
\]
On the bottom face $\mathcal{X}\times\{\delta\}$, we have
$w_\eta\ge 0$ by~\eqref{eq:bottom-face}. Moreover,
\[
\partial_t w_\eta-Lw_\eta
=
\partial_t v-Lv
\ge 0
\qquad\text{in }\mathcal{X}\times(\delta,T].
\]
By the weak minimum principle
(\cite[Theorem~6, Chapter~2, Section~2]{friedman2008partial}) applied to
$w_\eta$ on $\mathcal{X}\times(\delta,T]$,
\begin{equation}
\label{eq:w-eta}
w_\eta\ge 0
\qquad\text{on }\overline{\mathcal{X}}\times[\delta,T].
\end{equation}

\emph{Step 4: Limits.}
Since $\delta>0$ can be arbitrarily small,~\eqref{eq:w-eta} gives
\[
w_\eta\ge 0
\qquad\text{on } \overline{\mathcal{X}}\times(0,T].
\]
Letting $\eta\downarrow0$, we have $v\ge p$ on $\overline{\mathcal{X}}\times(0,T]$(Both $v$ and $p$ are continuous on
$\overline{\mathcal{X}}\times(0,T]$, so $v-p$ is continuous there.
Since $v-p\ge 0$ on the dense subset $\overline{\mathcal{X}}\times(0,T]$, it
follows that $v\ge p$ on $\overline{\mathcal{X}}\times(0,T]$).

\emph{Step 5: Evaluation at $t=T$.}
Since $\mathcal{X}_0\subset\overline{\mathcal{X}}$, we have
$p(\bm{x},T)\le v(\bm{x},T)\le\epsilon, \forall\bm{x}\in\mathcal{X}_0$.
But $p(\bm{x},T)=\mathbb{P}_{\mathrm{exit}}(\bm{x})$, so
\[
\mathbb{P}_{\mathrm{exit}}(\bm{x}_0)\le\epsilon,
\qquad
\forall\bm{x}_0\in\mathcal{X}_0,
\]
which is the desired conclusion.

\noindent\textbf{Part (ii): Necessity.}
Assume~\eqref{eq:strict-safety}. We construct a polynomial $v$
satisfying the strict version of~\eqref{eq:barrier}.

\emph{Step 1: Strict margin on $\mathcal{X}_0$.}
Since $\mathcal{X}_0$ is compact and $p(T,\cdot)$ is continuous on
$\mathcal{X}_0$ by~\eqref{eq:p-regularity}, there exists $\eta>0$ such
that
\begin{equation}
\label{eq:margin}
p(\bm{x},T)\le\epsilon-\eta,
\qquad
\forall\bm{x}\in\mathcal{X}_0.
\end{equation}

\emph{Step 2: Time-shifted candidate.}
Fix $\alpha>0$. For $\delta>0$, define
\begin{equation*}
\label{eq:v_delta}
v_\delta(\bm{x},t)
:=
p(\bm{x},t+\delta)+\delta e^{\alpha t},
\qquad
(\bm{x},t)\in\overline{\mathcal{X}}\times [0,T].
\end{equation*}
By~\eqref{eq:p-regularity} with $T$ replaced by $T+\delta$, we have
\begin{equation}
\label{eq:v_delta-regularity}
v_\delta\in C^{2+\alpha,1+\alpha/2}
\bigl(\overline{\mathcal{X}}\times [0,T]\bigr).
\end{equation}

\emph{Step 3: Verification of the strict barrier inequalities.}

\emph{(a) Supersolution.}
On $\overline{\mathcal{X}}\times [0,T]$,
\[
\partial_t v_\delta-Lv_\delta
=
\bigl(\partial_t p-Lp\bigr)(\cdot,\cdot+\delta)
+\alpha\delta e^{\alpha t}
=
0+\alpha\delta e^{\alpha t}
>0.
\]

\emph{(b) Lateral boundary.}
On $\partial\mathcal{X}\times[0,T]$,
$p(\cdot,\cdot+\delta)=1$, so
\[
v_\delta(\bm{x},t)
=
1+\delta e^{\alpha t}
>1.
\]

\emph{(c) Initial condition.}
On $\overline{\mathcal{X}}\times \{0\}$,
\[
v_\delta(\bm{x},0)
=
p(\bm{x},\delta)+\delta
\ge\delta
>0.
\]

\emph{(d) Terminal condition.}
Since $p$ is continuous on the compact set $[T,T+1]\times\overline{\mathcal{X}}$,
it is uniformly continuous there; hence
\[
\sup_{\bm{x}\in\overline{\mathcal{X}}}
\left|p(\bm{x},T+\delta)-p(\bm{x},T)\right|
\longrightarrow 0
\qquad\text{as }\delta\downarrow 0.
\]
Choose $\delta>0$ small enough that
\[
p(\bm{x},T+\delta)\le p(\bm{x},T)+\eta/2\le\epsilon-\eta/2,
\qquad
\forall\bm{x}\in\mathcal{X}_0,
\]
using~\eqref{eq:margin}. Decrease $\delta$ further if necessary so
that $\delta e^{\alpha T}<\eta/2$. Then
\[
v_\delta(\bm{x},T)=
p(\bm{x},T+\delta)+\delta e^{\alpha T}
<
\epsilon-\eta/2+\eta/2
=
\epsilon, 
\forall\bm{x}\in\mathcal{X}_0.
\]
Thus $v_\delta$ satisfies all four inequalities
in~\eqref{eq:barrier} with \emph{strict} sign.

\emph{Step 4: Polynomial approximation.}
By~\eqref{eq:v_delta-regularity}, a standard extension theorem for
Hölder spaces gives $\widetilde{v}_\delta\in C^{2,1}(U)$ for some
open neighbourhood $U\supset\overline{\mathcal{X}}\times [0,T]$, with
$\widetilde{v}_\delta=v_\delta$ on
$\overline{\mathcal{X}}\times [0,T]$.

Apply Lemma~\ref{lem:poly-approx} with
$K=\overline{\mathcal{X}}\times [0,T]$ and
$f=\widetilde{v}_\delta$. There is a sequence of polynomials
$\{v_k\}\subset\mathbb{R}[\bm{x},t]$ such that $\|v_k-v_\delta\|_{C^{2,1}(K)}\longrightarrow 0$.

\emph{Step 5: Strict inequalities survive approximation.}
Because each of the four inequalities holds strictly for
$v_\delta$ on a compact set, define the positive margins
\[
\begin{cases}
\eta_1
:=
\min_{\overline{\mathcal{X}}\times [0,T]}
\bigl(\partial_t v_\delta-Lv_\delta\bigr)>0,\\
\eta_2
:=
\min_{\partial\mathcal{X}\times[0,T]}
\bigl(v_\delta-1\bigr)>0,\\
\eta_3
:=
\min_{\overline{\mathcal{X}}}
v_\delta(\cdot,0)>0,\\
\eta_4
:=
\min_{\mathcal{X}_0}
\bigl(\epsilon-v_\delta(\cdot,T)\bigr)>0.
\end{cases}
\]
Let $\eta_*:=\tfrac12\min_i\eta_i>0$. Because $v_k\to v_\delta$
in $C^{2,1}(K)$, for all sufficiently large $k$,
\[
\|v_k-v_\delta\|_{C^{2,1}(K)}<\eta_*.
\]
Since $a_{ij},b_i$ are bounded and continuous on
$\overline{\mathcal{X}}$, the map $v\mapsto\partial_t v-Lv$ is
continuous with respect to the $C^{2,1}(K)$ norm, so
\[
\begin{split}
&\left\|
(\partial_t v_k-Lv_k)-(\partial_t v_\delta-Lv_\delta)
\right\|_{C(K)}\\
\le&
C_L\,\|v_k-v_\delta\|_{C^{2,1}(K)}< \eta_1/2.
\end{split}
\]
Hence, $\partial_t v_k-Lv_k \ge \eta_1/2>0$ on $\overline{\mathcal{X}}\times [0,T]$.
Similarly,
\[
\begin{cases}
v_k\ge 1+\eta_2/2,&\text{on }\partial\mathcal{X}\times[0,T],\\
v_k(\cdot,0)\ge\eta_3/2, &\text{on }\overline{\mathcal{X}},\\
v_k(\cdot,T)\le\epsilon-\eta_4/2, &\text{on }\mathcal{X}_0.
\end{cases}
\]
Thus, for all sufficiently large $k$, the polynomial $v_k$ satisfies the strict barrier condition~\eqref{eq:barrier}. This completes the proof of Part~(ii). \qed
\end{pf}

\section{Semidefinite Programming Computation}
\label{sec:sdp}

In this section, we synthesize polynomial barrier certificates for polynomial systems using SOS programming. We assume that the drift vector and the diffusion matrix have polynomial entries, and that the initial set, the safe set, and the target set are semialgebraic. Polynomial positivity conditions are replaced by membership in the corresponding Archimedean
quadratic modules, leading to a finite-dimensional SDP.
Putinar's Positivstellensatz guarantees that every strictly positive polynomial admits such a representation, which yields a complete degree hierarchy for polynomial barrier functions.

\subsection{Algebraic Representations}

\begin{assumption}
\label{ass:poly}
\begin{enumerate}
    \item The state-space regions appearing in the barrier-like
    condition~\eqref{eq:barrier} admit compact basic semialgebraic
    representations:
    \[
    \begin{cases}
        \overline{\mathcal{X}}
        = \bigl\{\bm{x}\in\mathbb{R}^n \mid
        g_{X,i}(\bm{x})\geq 0,\ i=1,\ldots,k_X\bigr\},\\[1ex]
        \partial\mathcal{X}
        = \bigl\{\bm{x}\in\mathbb{R}^n \mid
        g_{e,i}(\bm{x})\geq 0,\ i=1,\ldots,k_e\bigr\},\\[1ex]
        \mathcal{X}_0
        = \bigl\{\bm{x}\in\mathbb{R}^n \mid
        g_{0,i}(\bm{x})\geq 0,\ i=1,\ldots,k_0\bigr\},
    \end{cases}
    \]
    where $g_{X,i},g_{e,i},g_{0,i}\in\mathbb{R}[\bm{x}]$ are known
    polynomials.
    \item The components of $\bm{b}$ and the entries of $\bm{\sigma}$
    are polynomial in $\bm{x}$, and the barrier function $v$ is
    polynomial in $(\bm{x},t)$.
\end{enumerate}
\end{assumption}

Let
\[
\Sigma[\bm{x},t]:=
\left\{q\in\mathbb{R}[\bm{x},t]\middle|
\begin{split}
&q(\bm{x},t)=\sum_j h_j^2(\bm{x},t),\\
&h_j\in\mathbb{R}[\bm{x},t]
\end{split}
\right\}
\]
denote the cone of SOS polynomials in $(\bm{x},t)$. For a collection of
defining polynomials
$
G=\{g_1,\dots,g_k\}
$
in the variables $(\bm{x},t)$, the associated quadratic module is
\[\begin{split}
&\mathcal{M}(G):=\\
&\Bigl\{
\sigma_0(\bm{x},t)
+\sum_{i=1}^{k}\sigma_i(\bm{x},t)g_i(\bm{x},t)
\;\Big|\;
\sigma_0, \ldots,\sigma_k\in\Sigma[\bm{x},t]
\Bigr\}.
\end{split}
\]

To apply Putinar's Positivstellensatz we ensure that each quadratic
module is Archimedean. Since $\overline{\mathcal{X}}$ is compact, there
exists $R>0$ such that
\[
\overline{\mathcal{X}}\subseteq\{\bm{x}\mid R-\|\bm{x}\|_2^2\geq0\}.
\]
Define the redundant ball constraint
\[
g_{\mathrm{b}}(\bm{x}):=R-\|\bm{x}\|_2^2.
\]
Similarly, since $t\in[0,T]$, which can be equivalently formulated as $g_{t,T}(t)\geq 0$, where $g_{t,T}(t)=t(T-t)$, 
define the redundant time constraints
\[
g_{t,\mathrm{b}}(t):=T^2-t^2.
\]
The last constraint is redundant on $[0,T]$ and will be used only to make the quadratic modules Archimedean.

For the four regions in~\eqref{eq:barrier}, define the following
collections of defining polynomials:
\[
\begin{split}
&G_D=
\{g_{t,T},g_{t,\mathrm{b}},
g_{X,1},\ldots,g_{X,k_X},g_{\mathrm{b}}\},\\
&G_B=\{g_{t,T},g_{t,\mathrm{b}},
g_{e,1},\ldots,g_{e,k_e},g_{\mathrm{b}}\},\\
&G_I=\{g_{X,1},\ldots,g_{X,k_X},g_{\mathrm{b}}\},\\
&G_T=
\{g_{0,1},\ldots,g_{0,k_0},g_{\mathrm{b}}\}.
\end{split}
\]
Here $G_D$ corresponds to
$\overline{\mathcal{X}}\times[0,T]$, $G_B$ to
$\partial\mathcal{X}\times[0,T]$, $G_I$ to
$\overline{\mathcal{X}}$, and $G_T$ to
$\mathcal{X}_0$. The associated quadratic modules
\[\mathcal{M}(G_D),\mathcal{M}(G_B),\mathcal{M}(G_I),\mathcal{M}(G_T)\]
are Archimedean.

\begin{theorem}[Putinar's Positivstellensatz \cite{putinar1993positive}]
\label{thm:putinar}
Let $G=\{\bm{y} \in \mathbb{R}^m \mid g_i(\bm{y})\geq0,\ i=1,\ldots,k\}$ be a basic closed semialgebraic set. Suppose $\mathcal{M}(G)$ is Archimedean.
Then for every $p\in\mathbb{R}[\bm{y}]$ satisfying
$p(\bm{y})>0$ for all $\bm{y}\in G$, there exist SOS polynomials
$\sigma_0,\sigma_1,\ldots,\sigma_k\in\Sigma[\bm{y}]$ such that
$p(\bm{y})=\sigma_0(\bm{y})+\sum_{i=1}^{k}\sigma_i(\bm{y})g_i(\bm{y})$, i.e.\ $p\in\mathcal{M}(G)$.
\end{theorem}

Putinar's Positivstellensatz guarantees that, for an Archimedean basic closed semialgebraic set, every polynomial that is strictly positive on the set admits a representation in the corresponding quadratic module as a SOS with finite degree. Consequently, the SOS hierarchy is complete in the sense that every strictly positive polynomial constraint is certified at a finite hierarchy level. It has been widely applied in the reachability analysis literature, e.g.\
\cite{henrion2013convex,xue2019inner}. In the following, we show that the problem of
verifying the condition \eqref{eq:barrier} admits a SOS formulation that is both sound and complete, yielding a complete degree hierarchy for finite-horizon safety verification.

\subsection{SOS Reformulation and Soundness}

We now translate the four inequalities in the barrier-like
condition~\eqref{eq:barrier} into SOS constraints. Fix $\epsilon\in(0,1)$.
Define the four residual polynomials
\begin{equation}
\label{eq:residuals}
\begin{cases}
p_1(\bm{x},t):=
\frac{\partial v}{\partial t}(\bm{x},t)
-
Lv(\bm{x},t),\\[1ex]
p_2(\bm{x},t):=
v(\bm{x},t)-1,\\[1ex]
p_3(\bm{x}):=
v(\bm{x},0),\\[1ex]
p_4(\bm{x}):=
\epsilon-v(\bm{x},T).
\end{cases}
\end{equation}
We enforce the four inequalities by requiring each residual to belong to
the corresponding quadratic module:
\begin{equation*}
\label{putin_sos}
\begin{cases}
p_1(\bm{x},t)=
\sigma_{D,0}(\bm{x},t)
+\displaystyle\sum_{g\in G_D}\sigma_{D,g}(\bm{x},t)g(\bm{x},t),\\[1.2ex]
p_2(\bm{x},t)=
\sigma_{B,0}(\bm{x},t)
+\displaystyle\sum_{g\in G_B}\sigma_{B,g}(\bm{x},t)g(\bm{x},t),\\[1.2ex]
p_3(\bm{x})=
\sigma_{I,0}(\bm{x})
+\displaystyle\sum_{g\in G_I}\sigma_{I,g}(\bm{x})g(\bm{x}),\\[1.2ex]
p_4(\bm{x})=
\sigma_{T,0}(\bm{x})
+\displaystyle\sum_{g\in G_T}\sigma_{T,g}(\bm{x})g(\bm{x}),
\end{cases}
\end{equation*}
which is equivalent to the SOS constraints
\begin{equation}
\label{SOS}
\begin{cases}
p_1(\bm{x},t)
-\displaystyle\sum_{g\in G_D}\sigma_{D,g}(\bm{x},t)g(\bm{x},t)
\in\Sigma[\bm{x},t],\\[1.2ex]
p_2(\bm{x},t)
-\displaystyle\sum_{g\in G_B}\sigma_{B,g}(\bm{x},t)g(\bm{x},t)
\in\Sigma[\bm{x},t],\\[1.2ex]
p_3(\bm{x})
-\displaystyle\sum_{g\in G_I}\sigma_{I,g}(\bm{x})g(\bm{x})
\in\Sigma[\bm{x}],\\[1.2ex]
p_4(\bm{x})
-\displaystyle\sum_{g\in G_T}\sigma_{T,g}(\bm{x})g(\bm{x})
\in\Sigma[\bm{x}].
\end{cases}
\end{equation}
Here all multipliers are SOS polynomials in $(\bm{x},t)$.

\begin{theorem}[Soundness of the SOS certificate]
\label{thm:sos_soundness}
Suppose Assumptions~\ref{assump:sde_conditions},
\ref{assm2}, and~\ref{ass:poly} hold. Fix $\epsilon\in(0,1)$. If there
exist a polynomial $v\in\mathbb{R}[\bm{x},t]$ and SOS multipliers
satisfying~\eqref{SOS}, then $v$ satisfies the barrier-like
condition~\eqref{eq:barrier}, and therefore
\[
\mathbb{P}_{\mathrm{exit}}(\bm{x}_0)\le\epsilon,
\qquad
\forall\bm{x}_0\in\mathcal{X}_0.
\]
\end{theorem}
\begin{pf}
Every SOS polynomial is globally nonnegative, and every defining
polynomial is nonnegative on its associated semialgebraic set.
Therefore the right-hand side of each identity in~\eqref{putin_sos} is nonnegative on the corresponding set. Hence
\[
\begin{cases}
p_1\geq0 &\text{ on }\overline{\mathcal{X}}\times[0,T],\\
p_2\geq0 &\text{ on }\partial\mathcal{X}\times[0,T],\\
p_3\geq0 &\text{ on }\overline{\mathcal{X}},\\
p_4\geq0 &\text{ on }\mathcal{X}_0.
\end{cases}
\]
By definition of the residual polynomials~\eqref{eq:residuals}, this is exactly the barrier-like condition~\eqref{eq:barrier}. The guarantee $\mathbb{P}_{\mathrm{exit}}(\bm{x}_0)\le\epsilon$ on $\mathcal{X}_0$ then follows from the sufficiency direction of
Theorem~\ref{thm:barrier}.
\end{pf}

\subsection{Completeness of the SOS Hierarchy}

We now show that the SOS hierarchy is complete over polynomial barrier functions. This is a direct consequence of the necessity direction of Theorem~\ref{thm:barrier} combined with Putinar's Positivstellensatz.

\begin{theorem}[Completeness of the SOS hierarchy]
\label{cor:sos_complete}
Suppose Assumptions~\ref{assump:sde_conditions}--\ref{ass:poly} hold. If
\[
\mathbb{P}_{\mathrm{exit}}(\bm{x}_0)<\epsilon,
\qquad
\forall\bm{x}_0\in\mathcal{X}_0,
\]
then there exists a polynomial $v\in\mathbb{R}[\bm{x},t]$ such that the SOS program~\eqref{SOS} is feasible for some finite degree of the multipliers.
\end{theorem}
\begin{pf}
By the necessity direction of Theorem~\ref{thm:barrier}, there exists a
polynomial $v\in\mathbb{R}[\bm{x},t]$ such that the four barrier-like
inequalities in~\eqref{eq:barrier} hold strictly on the corresponding
compact regions. Consequently the residual
polynomials~\eqref{eq:residuals} satisfy
\[
p_1>0\text{ on }\overline{\mathcal{X}}\times[0,T],
\qquad
p_2>0\text{ on }\partial\mathcal{X}\times[0,T],
\]
\[
p_3>0\text{ on }\overline{\mathcal{X}},
\qquad
p_4>0\text{ on }\mathcal{X}_0.
\]
Since the four quadratic modules
\[\mathcal{M}(G_D),\mathcal{M}(G_B),\mathcal{M}(G_I),\mathcal{M}(G_T)\]
are Archimedean, Putinar's Positivstellensatz applies separately to
each region. Hence there exist SOS multipliers
\[
\left\{
\begin{split}
&\sigma_{D,0},\ldots,\sigma_{D,g},\ldots,
\sigma_{B,0},\ldots,\sigma_{B,g},\ldots,\\
&\sigma_{I,0},\ldots,\sigma_{I,g},\ldots,
\sigma_{T,0},\ldots,\sigma_{T,g},\ldots
\end{split}
\right\}
\]
such that~\eqref{putin_sos} holds, which is equivalent
to~\eqref{SOS}. Each representation involves finitely many SOS
polynomials and hence has finite degree. Thus, there exists a finite hierarchy level at which \eqref{SOS} is feasible. 

The proof is completed. \qed
\end{pf}

\begin{remark}
The completeness result is a statement about the existence of a polynomial certificate at some finite degree, not about the smallest such degree. The required degree is not known a priori. The degree hierarchy is obtained by increasing the degrees of the barrier polynomial and of the SOS multipliers until a feasible SDP is found.
\end{remark}
\section{Illustrative Examples}
\label{sec:ex}
In this section, we demonstrate the application of our theoretical developments through two numerical examples with polynomial systems. We aim to find polynomial barrier functions that satisfy the SOS constraints \eqref{SOS}. The resulting SOS program is a convex SDP in the coefficients of the polynomial $p$ and the Gram matrices of the SOS multipliers. In the experiments, the SOS program \eqref{SOS} is formulated using the MATLAB-based modeling toolbox YALMIP \cite{lofberg2004yalmip} and the resulting SDPs are solved using MOSEK 10.1.21 \cite{aps2019mosek}. To ensure numerical stability when solving these SDPs, we additionally impose a constraint on the coefficients of the unknown polynomials, specifically restricting them to the interval $[-10^2, 10^2]$. This restriction is imposed only in the numerical implementation and is not covered by the theoretical completeness guarantee.

\begin{example}
\label{ex1}
Consider the one-dimensional Brownian motion
\[
dX(t)=\sigma\,dW(t),
\qquad
X(t)\in\mathcal{X}:=(-1,1),
\]
with absorbing boundaries at $x=\pm1$. Here $b(x)=0$, $a(x)=\sigma^2$. Thus, the drift is polynomial and the diffusion matrix is uniformly elliptic.
The safe set $\mathcal{X}$ is bounded, open, and its boundary
$\partial\mathcal{X}=\{-1,1\}$ is a compact $C^\infty$ hypersurface.
Hence Assumptions \ref{assump:sde_conditions} and \ref{assm2} are satisfied.

The safety probability $u(x_0,T)=\mathbb{P}(\tau_{\mathcal{X}}>T)$
solves the parabolic initial-boundary value problem
\begin{equation}
\label{eq:benchmark-pde}
\begin{cases}
\partial_t u=\dfrac{\sigma^2}{2}\,\partial_{xx}u,
&(x,t)\in(-1,1)\times(0,T],\\[1mm]
u(\pm1,t)=0,
&t\in(0,T],\\[1mm]
\displaystyle\lim_{t\downarrow0}u(x,t)=1,
&x\in(-1,1).
\end{cases}
\end{equation}
By separation of variables, the unique classical solution of
\eqref{eq:benchmark-pde} admits the rapidly convergent eigenfunction
expansion
\begin{equation*}
\label{eq:benchmark-survival}
\begin{split}
u(x_0,T)=\frac{4}{\pi}
\sum_{k=0}^{\infty}
\frac{(-1)^k}{2k+1} a(k,x_0) b(k,x_0,T)
\end{split}
\end{equation*}
where $a(k,x_0)=\cos\!\left(\frac{(2k+1)\pi x_0}{2}\right)$ and $b(k,x_0,T)=\exp\!\left(-\frac{(2k+1)^2\pi^2\sigma^2T}{8}\right)$. The proof for this claim is shown in Appendix B. Consequently, the finite-horizon exit probability is
\begin{equation*}
\label{eq:benchmark-exit}
\begin{split}
\mathbb{P}_{\mathrm{exit}}(x_0)
=1-\frac{4}{\pi}
\sum_{k=0}^{\infty}
\frac{(-1)^k}{2k+1} a(k,x_0) b(k,x_0,T).
\end{split}
\end{equation*}

This benchmark is particularly attractive for validating the general theory because:
\begin{itemize}
\item the drift $b(x)=0$ is polynomial;
\item the diffusion $a(x)=\sigma^2$ is uniformly elliptic;
\item the safe set $\mathcal{X}=(-1,1)$ has a smooth boundary;
\item the exact solution is available as a rapidly convergent eigenfunction
expansion, enabling high-precision numerical validation of the general
proof.
\end{itemize}

Assume $\mathcal{X}_0=\{x\mid 0.01-x^2\geq 0\}$, $\sigma=1$, and $T=1$. According to the above analytical formula, we obtain 
\[\mathbb{P}_{\mathrm{exit}}(x_0)\leq 0.634, \forall \bm{x}_0\in \mathcal{X}_0.\]
Correspondingly, the computed feasibility results via solving \eqref{SOS} are reported in Table \ref{tab:sdpex1_set} with $g_{\mathrm{b}}=1-x^2$.
 
\begin{table}[H]
\caption{\centering Feasibility of SDP \eqref{SOS} for Example \ref{ex1}\\
(\ding{52}: feasible; \ding{55}: infeasible)}
\centering
\begin{tabular}{|c|c|c|c|c|}
\hline
Degree&$\epsilon=0.63$& $\epsilon=0.8$ & $\epsilon=0.7$ & $\epsilon=0.66$   
\\
\hline
4  &\ding{55} & \ding{52} & \ding{55} & \ding{55}  \\
6  &\ding{55}& \ding{52} & \ding{55} & \ding{55}  \\
8   &\ding{55}& \ding{52} & \ding{52} & \ding{55}  \\
10   &\ding{55}& \ding{52} & \ding{52} & \ding{55} \\
12   &\ding{55}& \ding{52} & \ding{52} & \ding{55} \\
14  & \ding{55}&\ding{52} & \ding{52} & \ding{55} \\
16  & \ding{55}&\ding{52} & \ding{52} & \ding{52} \\
\hline
\end{tabular}
\label{tab:sdpex1_set}
\end{table}

\noindent
\emph{Discussion.}
Table \ref{tab:sdpex1_set} exhibits the expected convergence of the SOS hierarchy: as the polynomial degree increases, progressively tighter bounds $\epsilon$ become feasible. At degree $4$ and $6$ only $\epsilon=0.8$ is certified, whereas at degree $16$ the SDP is feasible for $\epsilon=0.66$,
which is within $4\%$ of the analytical value
$\sup_{\mathcal X_0}\mathbb{P}_{\mathrm{exit}}(\tau_{\mathcal X}(x_0)\le 1)\approx 0.634$. The target $\epsilon=0.63$ remains infeasible at every degree considered, which is consistent with the analytical value: since the true supremum is  $0.634$, no polynomial certificate can certify a bound strictly below it, regardless of degree. More accurate upper bounds require polynomial barrier certificates of higher degree, which increases the computational burden.
\end{example}

\begin{example}
\label{ex2}
Consider the coupled SDE
\begin{equation*}
\label{eq:benchmark-coupled-sde}
\begin{cases}
dX_1(t)=
\bigl(-\theta X_1(t)-\omega X_2(t)\bigr)\,dt
+\sigma\,dW_1(t),\\
dX_2(t)=
\bigl(\omega X_1(t)-\theta X_2(t)\bigr)\,dt
+\sigma\,dW_2(t),
\end{cases}
\end{equation*}
on the disk $\mathcal{X}=\{x\in\mathbb R^2 \mid \|x\|<R\}$, with absorbing boundary.
The drift matrix
\[
A=
\begin{pmatrix}
-\theta&-\omega\\
\omega&-\theta
\end{pmatrix}
\]
couples the state variables whenever $\omega\neq0$. The safe set has
$C^\infty$ boundary, so Assumptions \ref{assump:sde_conditions}
and \ref{assm2} hold.

\noindent
\emph{Radial reduction.}
For a radial function $v(x)=\varphi(\|\bm{x}\|)$,
\[
L v
=
\frac{\sigma^2}{2}
\left(
\varphi''(r)+\frac1r\varphi'(r)
\right)
-\theta r\varphi'(r),
\]
where $r=\|\bm{x}\|$.
Thus $u(r,t)=\mathbb{P}(\tau_{\mathcal{X}}(\bm{x})>t)$ solves
\begin{equation*}
\label{eq:benchmark-coupled-radial}
\begin{cases}
\partial_t u
=
\dfrac{\sigma^2}{2}
\left(
u_{rr}+\dfrac1r u_r
\right)
-\theta r u_r,
&(r,t)\in(0,R)\times(0,T],\\[1mm]
u(R,t)=0,
\quad
u_r(0,t)=0,
&t\in(0,T],\\[1mm]
\displaystyle\lim_{t\downarrow0}u(r,t)=1,
&r\in(0,R).
\end{cases}
\end{equation*}
The finite-horizon exit probability is $\mathbb{P}_{\mathrm{exit}}(x_0)=1-u(\|x_0\|,T)$.

\noindent
\emph{Exact spectral solution.}
Set $z=\theta r^2/\sigma^2$. The eigenfunctions are
$\phi_n(r)={}_1F_1(-\alpha_n;1;z)$,
where $\alpha_n>0$ solves
\begin{equation}
\label{eq:benchmark-coupled-roots}
{}_1F_1\!\left(-\alpha_n;1;\frac{\theta R^2}{\sigma^2}\right)=0.
\end{equation}
where the function ${}_1F_1(a;b;z)$ is the \emph{confluent hypergeometric function}
of the first kind, also called \emph{Kummer's function}. It is defined by the
power series
\begin{equation}
\label{eq:kummer-series}
{}_1F_1(a;b;z)
=
\sum_{k=0}^{\infty}
\frac{(a)_k}{(b)_k}\frac{z^k}{k!},
\end{equation}
where $(a)_k$ is the Pochhammer symbol (rising factorial),
\[
(a)_0=1,
\qquad
(a)_k=a(a+1)(a+2)\cdots(a+k-1).
\]
Explicitly,
\begin{equation*}
\label{eq:kummer-explicit}
{}_1F_1(a;b;z)
=
1+\frac{a}{b}\frac{z}{1!}
+\frac{a(a+1)}{b(b+1)}\frac{z^2}{2!}
+\frac{a(a+1)(a+2)}{b(b+1)(b+2)}\frac{z^3}{3!}
+\cdots
\end{equation*}
The eigenvalues are $\lambda_n=2\theta\alpha_n$. Hence
\begin{equation*}
\label{eq:benchmark-coupled-exit}
\mathbb{P}_{\mathrm{exit}}(x_0)
=
1-
\sum_{n=1}^\infty
c_n\,
{}_1F_1\!\left(
-\alpha_n;1;
\frac{\theta\|x_0\|^2}{\sigma^2}
\right)
e^{-2\theta\alpha_nT},
\end{equation*}
with coefficients
\[
c_n
=
\frac{
\int_0^R
\phi_n(s)\,
s\,e^{-\theta s^2/\sigma^2}\,ds
}{
\int_0^R
\phi_n(s)^2\,
s\,e^{-\theta s^2/\sigma^2}\,ds
}.
\]
The proof for this claim is shown in Appendix B.

For $\theta=\sigma=R=\omega=1$, $\mathcal{X}_0=\{(x_1,x_2)^{\top}\mid 0.01-x_1^2-x_2^2\geq 0\}$, and $T=0.2$,
\[
\mathbb{P}_{\mathrm{exit}}(\bm{x}_0) \leq 0.15, \forall \bm{x}_0\in \mathcal{X}_0.
\]
Correspondingly, the feasibility results via solving \eqref{SOS} are reported in Table \ref{tab:sdpex2_set}.
\begin{table}[H]
\caption{\centering Feasibility of SDP \eqref{SOS} for Example \ref{ex2}\\
(\ding{52}: feasible; \ding{55}: infeasible)}
\centering
\begin{tabular}{|c|c|c|c|c|}
\hline
Degree& $\epsilon=0.14$ &$\epsilon=0.3$ & $\epsilon=0.2$  &$\epsilon=0.16$ 
\\
\hline
4  & \ding{55}  & \ding{52} & \ding{55} & \ding{55}  \\
6  & \ding{55} & \ding{52} & \ding{55} & \ding{55}  \\
8   & \ding{55} & \ding{52} & \ding{52} & \ding{55}  \\
10  & \ding{55}  & \ding{52} & \ding{52} & \ding{55} \\
12  & \ding{55}  & \ding{52} & \ding{52} & \ding{52} \\
\hline
\end{tabular}
\label{tab:sdpex2_set}
\end{table}

\noindent
\emph{Discussion.}
Table \ref{tab:sdpex2_set} exhibits the expected monotone convergence of the
SOS hierarchy with the one in Table \ref{tab:sdpex1_set} for Example \ref{ex1}. At degrees $4$ and $6$,
only $\epsilon=0.3$ is certified, while $\epsilon=0.2$ and $\epsilon=0.16$ remain infeasible. At degrees~$8$ and~$10$, the result tightens and $\epsilon=0.2$ becomes feasible, whereas $\epsilon=0.16$ is still infeasible. At degree $12$, the SDP becomes feasible for all three target values, including the tightest bound $\epsilon=0.16$. This monotone improvement with degree is consistent with the completeness statement of Theorem \ref{cor:sos_complete}: as the polynomial degree increases, the
Positivstellensatz certificate becomes richer and the certified bound approaches the true exit probability from above.
\end{example}

\emph{Computational complexity and choice of degree.}
The size of the SDP associated with \eqref{SOS} depends on the degrees of the barrier polynomial and SOS multipliers, as well as on the state dimension. A polynomial of total degree at most $d$ in the variables $(\bm{x},t)\in\mathbb{R}^{1+n}$ has
$$
\binom{n+d+1}{n+1}
$$
monomials. Thus, the number of monomials grows rapidly with the polynomial degree and state dimension. In Example \ref{ex1}, where $n=1$, this number is $\binom{d+2}{2}$, whereas in Example \ref{ex2}, where $n=2$, it is $\binom{d+3}{3}$. Moreover, the SOS multipliers introduce additional semidefinite constraints whose sizes depend on the corresponding monomial bases. Consequently, higher-degree relaxations can substantially increase the memory and computational requirements of the resulting SDP.

Although higher-degree relaxations may yield tighter numerical upper bounds, we do not consider them in our experiments because of their additional computational cost for the examples studied. The degree-$16$ relaxation reported in Table \ref{tab:sdpex1_set} and the degree-$12$ relaxation reported in Table \ref{tab:sdpex2_set} therefore represent practical choices for balancing bound tightness against computational cost.

\section{Conclusion}
\label{sec:conclusion}

This paper developed a novel theoretical framework for finite-horizon safety verification of stochastic continuous-time systems through barrier certificates under uniform ellipticity of the diffusion matrix and regularity of the safe-set boundary.  Starting from the PDE characterization of the safety probability, we formulated barrier-like inequalities as an analytical relaxation of the PDE characterization. We proved that this relaxation is sound and complete in the sense of barrier certificates: every feasible barrier certifies implies safety, and, conversely, whenever the safety property holds with a strict margin, a polynomial barrier certificate exists. Furthermore, for polynomial systems over semialgebraic domains, we showed that polynomial barrier certificates can be synthesized through SOS programming. This establishes a connection between finite-horizon safety verification and computational certificate synthesis while preserving soundness and completeness guarantees.

Future work will investigate scalable algorithms for synthesizing polynomial barrier certificates and extensions to more general stochastic and learning-enabled systems.

\section*{ACKNOWLEDGMENT}
OpenAI’s GPT-5.6 Luna (free version) \cite{chang2026chatgpt} and Deepseek (free version) \cite{liu2024deepseek} were primarily used to polish the language of this paper and to provide feedback prior to submission. No methods or computational results presented in this work were generated by AI models. The authors take full responsibility for the accuracy and integrity of all content reported herein.


\bibliographystyle{unsrt}
\bibliography{ref}

@article{xue2026sufficient,
  title={Sufficient and necessary barrier-like conditions for safety and reach-avoid verification of stochastic discrete-time systems},
  author={Xue, Bai},
  journal={Automatica},
  volume={187},
  pages={112919},
  year={2026},
  publisher={Elsevier}
}

@article{cao2025comparative,
  title={Comparative analysis of barrier-like function methods for reach-avoid verification in stochastic discrete-time systems},
  author={Cao, Zhipeng and Wang, Peixin and Ong, Luke and {\v{Z}}ikeli{\'c}, {\DJ}or{\dj}e and Wagner, Dominik and Xue, Bai},
  journal={arXiv preprint arXiv:2512.05348},
  year={2025}
}

@article{xue2026sufficientc,
  title={Sufficient and Necessary Smooth Barrier-like Conditions for Continuous-Time Stochastic Reach-Avoid Verification},
  author={Xue, Bai},
  journal={arXiv preprint arXiv:2609.23981},
  year={2026}
}

@article{xue2026sufficientd,
  title={Sufficient and Necessary Continuous Barrier-like Conditions for Discrete-Time Stochastic Reach-Avoid Verification},
  author={Xue, Bai},
  journal={arXiv preprint arXiv:2609.23973},
  year={2026}
}

@inproceedings{yu2023safe,
  title={Safe probabilistic invariance verification for stochastic discrete-time dynamical systems},
  author={Yu, Yiqing and Wu, Taoran and Xia, Bican and Wang, Ji and Xue, Bai},
  booktitle={2023 62nd IEEE Conference on Decision and Control (CDC)},
  pages={5804--5811},
  year={2023},
  organization={IEEE}
}

@inproceedings{peruffo2021automated,
  title={Automated and formal synthesis of neural barrier certificates for dynamical models},
  author={Peruffo, Andrea and Ahmed, Daniele and Abate, Alessandro},
  booktitle={International conference on tools and algorithms for the construction and analysis of systems},
  pages={370--388},
  year={2021},
  organization={Springer}
}

@article{kordabad2026certificates,
  title={On certificates for almost sure reachability in stochastic systems},
  author={Kordabad, Arash Bahari and Majumdar, Rupak and Motwani, Harshit Jitendra and Soudjani, Sadegh},
  journal={IEEE Transactions on Automatic Control},
  year={2026},
  publisher={IEEE}
}

@article{parrilo2003semidefinite,
  title={Semidefinite programming relaxations for semialgebraic problems},
  author={Parrilo, Pablo A},
  journal={Mathematical programming},
  volume={96},
  number={2},
  pages={293--320},
  year={2003},
  publisher={Springer}
}

@inproceedings{abate2024stochastic,
  title={Stochastic omega-regular verification and control with supermartingales},
  author={Abate, Alessandro and Giacobbe, Mirco and Roy, Diptarko},
  booktitle={International Conference on Computer Aided Verification},
  pages={395--419},
  year={2024},
  organization={Springer}
}

@article{xue2026converse,
  title={Converse Barrier Certificates for Set-Based Stochastic Reach-Avoid Verification},
  author={Xue, Bai and Ong, C-H Luke},
  journal={arXiv preprint arXiv:2608.30318},
  year={2026}
}

@article{li2026converse,
  title={Converse barrier certificates for finite-time safety verification of continuous-time perturbed deterministic systems},
  author={Li, Yonghan and Wu, Chenyu and Wu, Taoran and Wang, Shijie and Xue, Bai},
  journal={Systems \& Control Letters},
  volume={209},
  pages={106357},
  year={2026},
  publisher={Elsevier}
}

@inproceedings{lofberg2004yalmip,
  title={YALMIP: A toolbox for modeling and optimization in MATLAB},
  author={Lofberg, Johan},
  booktitle={2004 IEEE international conference on robotics and automation (IEEE Cat. No. 04CH37508)},
  pages={284--289},
  year={2004},
  organization={IEEE}
}

@inproceedings{liu2019reachability,
  title={Reachability estimation of stochastic dynamical systems by semi-definite programming},
  author={Liu, Kairong and Li, Meilun and She, Zhikun},
  booktitle={2019 IEEE 58th Conference on Decision and Control (CDC)},
  pages={7727--7732},
  year={2019},
  organization={IEEE}
}

@inproceedings{bujorianu2007new,
  title={New insights on stochastic reachability},
  author={Bujorianu, Manuela L and Lygeros, John},
  booktitle={2007 46th IEEE Conference on Decision and Control},
  pages={6172--6177},
  year={2007},
  organization={IEEE}
}

@inproceedings{vzikelic2023learning,
  title={Learning control policies for stochastic systems with reach-avoid guarantees},
  author={{\v{Z}}ikeli{\'c}, {\DJ}or{\dj}e and Lechner, Mathias and Henzinger, Thomas A and Chatterjee, Krishnendu},
  booktitle={Proceedings of the AAAI Conference on Artificial Intelligence},
  volume={37},
  number={10},
  pages={11926--11935},
  year={2023}
}

@article{santoyo2021barrier,
  title={A barrier function approach to finite-time stochastic system verification and control},
  author={Santoyo, Cesar and Dutreix, Maxence and Coogan, Samuel},
  journal={Automatica},
  volume={125},
  pages={109439},
  year={2021},
  publisher={Elsevier}
}

@article{aps2019mosek,
  title={Mosek optimization toolbox for matlab},
  author={ApS, Mosek},
  journal={User’s Guide and Reference Manual, Version},
  volume={4},
  number={1},
  pages={116},
  year={2019}
}

@article{chen2025construction,
  title={On the Construction of Barrier Certificate: A Dynamic Programming Perspective},
  author={Chen, Yu and Li, Shaoyuan and Yin, Xiang},
  journal={arXiv preprint arXiv:2507.17222},
  year={2025}
}

@article{xue2024sufficient,
   title={Sufficient and necessary barrier-like conditions for safety and reach-avoid verification of stochastic discrete-time systems},
  author={Xue, Bai},
  journal={Automatica},
  volume={187},
  pages={112919},
  year={2026},
  publisher={Elsevier}
}

@book{oksendal2013stochastic,
  title={Stochastic differential equations: an introduction with applications},
  author={Oksendal, Bernt},
  year={2013},
  publisher={Springer Science \& Business Media}
}

@inproceedings{prajna2004safety,
  title={Safety verification of hybrid systems using barrier certificates},
  author={Prajna, Stephen and Jadbabaie, Ali},
  booktitle={HSCC'04},
  pages={477--492},
  year={2004},
  organization={Springer}
}

@article{prajna2007convex,
  title={Convex programs for temporal verification of nonlinear dynamical systems},
  author={Prajna, Stephen and Rantzer, Anders},
  journal={SIAM J. Control Optim.},
  volume={46},
  number={3},
  pages={999--1021},
  year={2007},
  publisher={SIAM}
}

@book{ville1939etude,
  title={Etude critique de la notion de collectif},
  author={Ville, Jean},
  year={1939},
  publisher={Gauthier-Villars Paris}
}

@article{xue2023new,
  title={A new framework for bounding reachability probabilities of continuous-time stochastic systems},
  author={Xue, Bai},
  journal={Nonlinear Analysis: Hybrid Systems},
  volume={60},
  pages={101670},
  year={2026},
  publisher={Elsevier}
}

@article{xue2019inner,
  title={Inner-approximating reachable sets for polynomial systems with time-varying uncertainties},
  author={Xue, Bai and Fr{\"a}nzle, Martin and Zhan, Naijun},
  journal={IEEE Transactions on Automatic Control},
  volume={65},
  number={4},
  pages={1468--1483},
  year={2019},
  publisher={IEEE}
}

@inproceedings{jagtap2018temporal,
  title={Temporal logic verification of stochastic systems using barrier certificates},
  author={Jagtap, Pushpak and Soudjani, Sadegh and Zamani, Majid},
  booktitle={International Symposium on Automated Technology for Verification and Analysis},
  pages={177--193},
  year={2018},
  organization={Springer}
}

@article{nachbin1949algebres,
  title={Sur les alg{\`e}bres denses de fonctions diff{\'e}rentiables sur une vari{\'e}t{\'e}},
  author={Nachbin, Leopoldo},
  journal={CR Acad. Sci. Paris},
  volume={228},
  pages={1549--1551},
  year={1949}
}

@article{xue2025refined,
  title={Refined Barrier Conditions for Finite-Time Safety and Reach-Avoid Guarantees in Stochastic Systems},
  author={Xue, Bai and Ong, Luke and Wagner, Dominik and Wang, Peixin},
  journal={arXiv preprint arXiv:2509.18518},
  year={2025}
}

@inproceedings{papachristodoulou2005tutorial,
  title={A tutorial on sum of squares techniques for systems analysis},
  author={Papachristodoulou, Antonis and Prajna, Stephen},
  booktitle={Proceedings of the 2005, American Control Conference, 2005.},
  pages={2686--2700},
  year={2005},
  organization={IEEE}
}

@inproceedings{wang2021safety,
  title={Safety-critical control of stochastic systems using stochastic control barrier functions},
  author={Wang, Chuanzheng and Meng, Yiming and Smith, Stephen L and Liu, Jun},
  booktitle={2021 60th IEEE Conference on Decision and Control (CDC)},
  pages={5924--5931},
  year={2021},
  organization={IEEE}
}

@article{steinhardt2012finite,
  title={Finite-time regional verification of stochastic non-linear systems},
  author={Steinhardt, Jacob and Tedrake, Russ},
  journal={The International Journal of Robotics Research},
  volume={31},
  number={7},
  pages={901--923},
  year={2012},
  publisher={SAGE Publications Sage UK: London, England}
}

@inproceedings{bujorianu2009stochastic,
  title={Stochastic reachability as an exit problem},
  author={Bujorianu, Manuela L and Blom, Henk AP},
  booktitle={2009 17th Mediterranean Conference on Control and Automation},
  pages={1026--1031},
  year={2009},
  organization={IEEE}
}

@inproceedings{xue2021reach,
  title={Reach-avoid analysis for stochastic discrete-time systems},
  author={Xue, Bai and Li, Renjue and Zhan, Naijun and Fr{\"a}nzle, Martin},
  booktitle={2021 American Control Conference (ACC)},
  pages={4879--4885},
  year={2021},
  organization={IEEE}
}

@article{esfahani2016stochastic,
  title={The stochastic reach-avoid problem and set characterization for diffusions},
  author={Esfahani, Peyman Mohajerin and Chatterjee, Debasish and Lygeros, John},
  journal={Automatica},
  volume={70},
  pages={43--56},
  year={2016},
  publisher={Elsevier}
}

@article{prajna2007framework,
  title={A framework for worst-case and stochastic safety verification using barrier certificates},
  author={Prajna, Stephen and Jadbabaie, Ali and Pappas, George J},
  journal={IEEE Transactions on Automatic Control},
  volume={52},
  number={8},
  pages={1415--1428},
  year={2007},
  publisher={IEEE}
}

@article{koutsoukos2008computational,
  title={Computational methods for verification of stochastic hybrid systems},
  author={Koutsoukos, Xenofon D and Riley, Derek},
  journal={IEEE Transactions on Systems, Man, and Cybernetics-Part A: Systems and Humans},
  volume={38},
  number={2},
  pages={385--396},
  year={2008},
  publisher={IEEE}
}

@inproceedings{prajna2004stochastic,
  title={Stochastic safety verification using barrier certificates},
  author={Prajna, Stephen and Jadbabaie, Ali and Pappas, George J},
  booktitle={2004 43rd IEEE conference on decision and control (CDC)(IEEE Cat. No. 04CH37601)},
  volume={1},
  pages={929--934},
  year={2004},
  organization={IEEE}
}

@article{henrion2013convex,
  title={Convex computation of the region of attraction of polynomial control systems},
  author={Henrion, Didier and Korda, Milan},
  journal={IEEE Transactions on Automatic Control},
  volume={59},
  number={2},
  pages={297--312},
  year={2013},
  publisher={IEEE}
}

@article{majumdar2025sound,
  title={Sound and complete proof rules for probabilistic termination},
  author={Majumdar, Rupak and Sathiyanarayana, VR},
  journal={Proceedings of the ACM on Programming Languages},
  volume={9},
  number={POPL},
  pages={1871--1902},
  year={2025},
  publisher={ACM New York, NY, USA}
}

@inproceedings{feng2020unbounded,
  title={Unbounded-time safety verification of stochastic differential dynamics},
  author={Feng, Shenghua and Chen, Mingshuai and Xue, Bai and Sankaranarayanan, Sriram and Zhan, Naijun},
  booktitle={International Conference on Computer Aided Verification},
  pages={327--348},
  year={2020},
  organization={Springer}
}

@article{liu2024deepseek,
  title={Deepseek-v3 technical report},
  author={Liu, Aixin and Feng, Bei and Xue, Bing and Wang, Bingxuan and Wu, Bochao and Lu, Chengda and Zhao, Chenggang and Deng, Chengqi and Zhang, Chenyu and Ruan, Chong and others},
  journal={arXiv preprint arXiv:2412.19437},
  year={2024}
}

@article{maghenem2022converse,
  title={On the converse safety problem for differential inclusions: Solutions, regularity, and time-varying barrier functions},
  author={Maghenem, Mohamed and Sanfelice, Ricardo G},
  journal={IEEE Transactions on Automatic Control},
  volume={68},
  number={1},
  pages={172--187},
  year={2022},
  publisher={IEEE}
}

@article{meng2022smooth,
  title={Smooth converse Lyapunov-barrier theorems for asymptotic stability with safety constraints and reach-avoid-stay specifications},
  author={Meng, Yiming and Li, Yinan and Fitzsimmons, Maxwell and Liu, Jun},
  journal={Automatica},
  volume={144},
  pages={110478},
  year={2022},
  publisher={Elsevier}
}

@inproceedings{kordabad2024control,
  title={Control barrier functions for stochastic systems under signal temporal logic tasks},
  author={Kordabad, Arash Bahari and Charitidou, Maria and Dimarogonas, Dimos V and Soudjani, Sadegh},
  booktitle={2024 European Control Conference (ECC)},
  pages={3213--3219},
  year={2024},
  organization={IEEE}
}

@article{chang2026chatgpt,
  title={ChatGPT: https://chatgpt. com},
  author={Chang, Qizhong},
  journal={RELC Journal},
  volume={57},
  number={1},
  pages={269--275},
  year={2026},
  publisher={SAGE Publications Sage UK: London, England}
}

@article{laurenti2025unifying,
  title={A unifying perspective for safety of stochastic systems: From barrier functions to finite abstractions},
  author={Laurenti, Luca and Lahijanian, Morteza},
  journal={IEEE Transactions on Automatic Control},
  year={2025},
  publisher={IEEE}
}

@article{putinar1993positive,
  title={Positive polynomials on compact semi-algebraic sets},
  author={Putinar, Mihai},
  journal={Indiana University Mathematics Journal},
  volume={42},
  number={3},
  pages={969--984},
  year={1993},
  publisher={JSTOR}
}

@article{xue2024,
  author={Xue, Bai and Zhan, Naijun and Fränzle, Martin},
  journal={IEEE Transactions on Automatic Control}, 
  title={Reach-Avoid Analysis for Polynomial Stochastic Differential Equations}, 
  year={2024},
  volume={69},
  number={3},
  pages={1882-1889}
}

@article{shreve1991brownian,
  title={Brownian motion and stochastic calculus},
  author={Karatzas, Ioannis and Shreve, Steven E},
  journal={Graduate texts in mathematics},
  volume={113},
  year={1991}
}

@book{friedman2008partial,
  title={Partial differential equations of parabolic type},
  author={Friedman, Avner},
  year={2008},
  publisher={Courier Dover Publications}
}

@article{jagtap2020formal,
  title={Formal synthesis of stochastic systems via control barrier certificates},
  author={Jagtap, Pushpak and Soudjani, Sadegh and Zamani, Majid},
  journal={IEEE Transactions on Automatic Control},
  volume={66},
  number={7},
  pages={3097--3110},
  year={2020},
  publisher={IEEE}
}

\newpage
\section*{Appendix A}
\label{A}
\setcounter{lemma}{0}
\begin{lemma}[Short-time exit probability]
\label{lem:short-time-exit}
For $\rho>0$, set $K:=\bigl\{\bm{x}\in\mathcal X \mid \operatorname{dist}(\bm{x},\partial\mathcal X)\ge\rho\bigr\}$. Then there exists $\delta_0>0$ such that for all $0<\delta\le\delta_0$,
\begin{equation}
\label{eq:short-time-bound}
\mathbb{P}(\tau_{\mathcal X}\le\delta)
\le
2n\exp\!\left(-\frac{\rho^2}{8nS^2\delta}\right),
\qquad
\forall\bm{x}\in K,
\end{equation}
where $S:=\sup_{\bm z\in\overline{\mathcal X}}\|\sigma(\bm z)\|_F<\infty$. Consequently,
\[
\lim_{\delta\downarrow0}\sup_{\bm{x}\in K}\mathbb{P}(\tau_{\mathcal X}\le\delta)=0.
\]
\end{lemma}
\begin{pf}
Fix $\bm{x}\in K$. Define the stopped vector martingale $\bm M_t:=\int_0^{t\wedge\tau_{\mathcal X}}\sigma(\bm{X}_{\bm{x}}(s))\,d\bm W_s$, whose $i$-th coordinate is $M_t^{(i)}=e_i^\top\bm M_t$.
The stopped SDE gives
\begin{equation}
\label{eq:stopped-sde}
\bm{X}_{\bm{x}}(t\wedge\tau_{\mathcal X})-\bm{x}
=
\int_0^{t\wedge\tau_{\mathcal X}}\bm b(\bm{X}_{\bm{x}}(s))\,ds+\bm M_t.
\end{equation}

\noindent\emph{Step 1: Reduction to a martingale event (case $S>0$).}
On $\{\tau_{\mathcal X}\le\delta\}$, continuity of the sample paths
gives $\bm{X}_{\bm{x}}(\tau_{\mathcal X})\in\partial\mathcal X$. Hence, by
\eqref{eq:stopped-sde} and the reverse triangle inequality,
\[
\begin{split}
\|\bm M_{\tau_{\mathcal X}}\|
&\ge
\|\bm{X}_{\bm{x}}(\tau_{\mathcal X})-\bm{x}\|
-
\left\|\int_0^{\tau_{\mathcal X}}\bm b(\bm{X}_{\bm{x}}(s))\,ds\right\|\\
&\ge \rho-B\tau_{\mathcal X}
\ge \rho-B\delta,
\end{split}
\]
where $B:=\sup_{\bm z\in\overline{\mathcal X}}\|\bm b(\bm z)\|<\infty$. Choose $\delta$ small enough that $B\delta\le\rho/2$ (if $B=0$,
this restriction is unnecessary). Then $\{\tau_{\mathcal X}\le\delta\}
\subseteq
\Bigl\{\sup_{0\le t\le\delta}\|\bm M_t\|\ge\frac{\rho}{2}\Bigr\}$,
and consequently
\[
\mathbb{P}(\tau_{\mathcal X}\le\delta)
\le
\mathbb{P}\!\left(
\sup_{0\le t\le\delta}\|\bm M_t\|\ge\frac{\rho}{2}
\right).
\]

\noindent\emph{Step 2: Exponential maximal inequality.}
Fix $i\in\{1,\dots,n\}$ and $\lambda>0$. Since $M_t^{(i)}$ is a
continuous local martingale, its quadratic variation $[M^{(i)}]_t$
coincides with the predictable quadratic variation $\langle M^{(i)}\rangle_t$;
see \cite[Theorem~4.10 and Definition~5.3]{shreve1991brownian}. Moreover,
the It\^o isometry gives
\begin{equation*}
\label{eq:qv}
[M^{(i)}]_t
=
\langle M^{(i)}\rangle_t
=
\int_0^{t\wedge\tau_{\mathcal X}}
\|e_i^\top\sigma(\bm{X}_{\bm{x}}(s))\|^2\,ds
\le
S^2 t.
\end{equation*}

Define the continuous semimartingale
\[
Y_t= \lambda M_t^{(i)}+A_t,
\qquad
A_t
:=
-\frac{\lambda^2}{2}\langle M^{(i)}\rangle_t.
\]
Since $\langle M^{(i)}\rangle_t$ is continuous, non-decreasing, and
finite-valued, the process $A_t$ is of finite variation. Therefore
its quadratic variation vanishes, $[A]_t=0$, and its quadratic covariation with the continuous local martingale
$M_t^{(i)}$ also vanishes, $[M^{(i)},A]_t=0$. Since the processes are continuous, the predictable quadratic
covariations coincide with the quadratic covariations, so
$\langle A\rangle_t=0$ and $\langle M^{(i)},A\rangle_t=0$.
Consequently, $d\langle Y\rangle_t= \lambda^2\,d\langle M^{(i)}\rangle_t$.
i.e.\ the finite‑variation part $A_t$ contributes nothing to the
quadratic variation of $Y_t$.

By It\^o's formula applied to $Z_t=\exp(Y_t)$,
\[
dZ_t
=
Z_t\,dY_t
+
\frac{1}{2}Z_t\,d\langle Y\rangle_t.
\]
Substituting $dY_t=\lambda\,dM_t^{(i)}-\frac{\lambda^2}{2}\,d\langle M^{(i)}\rangle_t$ and $d\langle Y\rangle_t=\lambda^2\,d\langle M^{(i)}\rangle_t$, we
obtain
\[
\begin{aligned}
dZ_t
&=
Z_t\!\left(
\lambda\,dM_t^{(i)}
-
\frac{\lambda^2}{2}\,d\langle M^{(i)}\rangle_t
\right)
+
\frac{1}{2}Z_t\,\lambda^2\,d\langle M^{(i)}\rangle_t\\
&=
\lambda Z_t\,dM_t^{(i)}.
\end{aligned}
\]
 Thus $Z_t$ is a positive local martingale, because it is a stochastic integral with respect to the continuous local martingale $M_t^{(i)}$ with locally bounded integrand $\lambda Z_t$. Hence $Z_t$ is a supermartingale. Since $Z_0=\exp(0)=1$, we have $\mathbb E[Z_0]=1$.
On the event $\{\sup_{0\le t\le\delta}M_t^{(i)}\ge r\}$,
\[
\sup_{0\le t\le\delta}Z_t
\ge
\exp\!\left(\lambda r-\frac{\lambda^2 S^2\delta}{2}\right),
\]
so Ville's inequality for nonnegative supermartingales \cite{ville1939etude} yields $\mathbb{P}\!\left(\sup_{0\le t\le\delta}M_t^{(i)}\ge r\right)
\le
\exp\!\left(-\lambda r+\frac{\lambda^2 S^2\delta}{2}\right)$.

Optimizing in $\lambda>0$ gives $\lambda=r/(S^2\delta)$ ($S>0$
from Assumption~\ref{assm2}), so $\mathbb{P}\!\left(\sup_{0\le t\le\delta}M_t^{(i)}\ge r\right) \le \exp\!\left(-\frac{r^2}{2S^2\delta}\right)$.

Applying the same argument to $-M_t^{(i)}$ and combining yields
the two-sided bound
\[
\mathbb{P}\!\left(\sup_{0\le t\le\delta}|M_t^{(i)}|\ge r\right)
\le
2\exp\!\left(-\frac{r^2}{2S^2\delta}\right).
\]

\noindent\emph{Step 3: Union bound.}
If $\|\bm M_t\|\ge\rho/2$, then at least one coordinate satisfies
$|M_t^{(i)}|\ge\rho/(2\sqrt n)$. Hence, by the union bound,
\[
\begin{aligned}
\mathbb{P}(\tau_{\mathcal X}\le\delta)
&\le
\sum_{i=1}^n
\mathbb{P}\!\left(
\sup_{0\le t\le\delta}|M_t^{(i)}|
\ge\frac{\rho}{2\sqrt n}
\right)\\
&\le
2n\exp\!\left(-\frac{\rho^2}{8nS^2\delta}\right).
\end{aligned}
\]
This bound is independent of $\bm{x}\in K$, so
\eqref{eq:short-time-bound} holds uniformly on $K$.

\noindent
Combining Steps~1--3, \eqref{eq:short-time-bound} holds for all
sufficiently small $\delta>0$, and letting $\delta\downarrow0$
gives the stated limit. \qed
\end{pf}

\begin{lemma}
\label{lemma:boundary}
Let $\tau_{\mathcal{X}}(\bm{x}_0) \equiv \inf \{ t \ge 0 \mid \bm{X}_{\bm{x}_0}(t) \notin \mathcal{X} \}$ be the first exit time, and define the exit probability before the time instant $t$ by:
\[
\mathbb{P}_{\mathrm{exit}}(\bm{x}, t) := \mathbb{P}(\tau_{\mathcal{X}}(\bm{x}) \le t), \quad \forall (\bm{x}, t) \in \overline{\mathcal{X}} \times [0, \hat{T}],
\]
where $\hat{T}>0$ is an arbitrary but fixed scalar value.

Under Assumptions \ref{assump:sde_conditions} and \ref{assm2}, the following conclusions hold: 
\begin{enumerate}
    \item[(i)] The safety probability $u(\bm{x},t) = 1-\mathbb{P}_{\mathrm{exit}}(\bm{x}, t)$ is the unique bounded classical solution in 
\[
u\in C^{2,1}\bigl(\mathcal{X}\times(0,\hat T]\bigr)
\cap
C\bigl(\overline{\mathcal{X}}\times(0,\hat T]\bigr)\cap C\bigl(\mathcal{X}\times[0,\hat T]\bigr),
\]
to the parabolic initial-boundary value problem:
\begin{equation}
\label{eq:survival_pde}
\begin{cases}
\partial_t u(\bm{x},t)=Lu(\bm{x},t),
&(\bm{x},t)\in\mathcal{X}\times(0,\hat T],\\
u(\bm{x},t)=0,
&(\bm{x},t)\in\partial\mathcal{X}\times(0,\hat T],\\
\displaystyle\lim_{t\downarrow0}u(\bm{x},t)=1,
&\bm{x}\in\mathcal{X},\\
u(\bm{x},0)=1, & \bm{x}\in \mathcal{X}.
\end{cases}
\end{equation}
\item[(ii)] For every $\delta\in(0,\hat T)$,
\[
u\in
C^{2+\alpha,\,1+\alpha/2}
\bigl(\overline{\mathcal{X}}\times[\delta,\hat T]\bigr)
\]
where $\alpha\in (0,\gamma]$ and $L$ denotes the spatial infinitesimal generator
\[
Lv(\bm{x},t)
=
\nabla_{\bm{x}}v(\bm{x},t)^\top\bm b(\bm{x})
+
\frac12
\operatorname{tr}\!\bigl(\bm a(\bm{x})\bm H_{\bm{x}}v(\bm{x},t)\bigr).
\]
\end{enumerate}
\end{lemma}
\begin{pf}
We first prove the conclusion in \textbf{Part i}.

\textbf{Step 1: Approximation of the initial datum.}

Choose
\[
\tilde d\in C^{2,\gamma}(\overline{\mathcal{X}}),\qquad
\tilde d>0\ \text{in }\mathcal{X},\qquad
\tilde d=0\ \text{on }\partial\mathcal{X}.
\]
Let $\eta\in C^\infty([0,\infty))$ satisfy
\[
0\le\eta\le1,\quad
\eta(r)=0\ \text{for }r\le1,\quad
\eta(r)=1\ \text{for }r\ge2,
\]
and define
\[
\phi_m(\bm{x}):=\eta\bigl(m\tilde d(\bm{x})\bigr).
\]
Then
\[
\phi_m\in C^{2+\gamma}(\overline{\mathcal{X}})
\subset C^{2+\alpha}(\overline{\mathcal{X}}),\qquad
0\le\phi_m\le1,
\]
and
\[
\phi_m=0\ \text{on }\partial\mathcal{X},\qquad
\phi_m\to1\ \text{locally uniformly in }\mathcal{X}.
\]
Since $\tilde d=0$ on $\partial\mathcal{X}$ and $\eta(r)=0$ for $r\le1$, $\phi_m$ vanishes in the whole neighborhood
\[
\{\bm{x}\in\mathcal{X} \mid \tilde d(\bm{x})\le 1/m\}
\]
of $\partial\mathcal{X}$. Hence all spatial derivatives of $\phi_m$ vanish on $\partial\mathcal{X}$. In particular,
\[
L\phi_m=0\qquad\text{on }\partial\mathcal{X}.
\]

Since $\phi_m$ vanishes in a neighborhood of
$\partial\mathcal X$, its spatial derivatives up to order two
vanish on $\partial\mathcal X$. Consequently,
\[
\phi_m=0,\qquad L\phi_m=0
\quad\text{on }\partial\mathcal X.
\]
Thus the compatibility conditions required for the
$C^{2+\alpha,1+\alpha/2}$ regularity of the approximating
solution are satisfied.

\noindent
\textbf{Step 2: Classical approximating problems.}

Define the function $\psi_m$ on the closed cylinder by
\begin{equation}
\label{eq:psi_m}
\psi_m(\bm{x},t):=\phi_m(\bm{x}),
\qquad
(\bm{x},t)\in\overline{\mathcal{X}}\times[0,\hat{T}].
\end{equation}
Since $\phi_m=0$ on $\partial\mathcal{X}$, this constant-in-time
extension matches the prescribed initial and boundary data. Moreover,
$\phi_m\in C^{2+\gamma}(\overline{\mathcal{X}})\subset
C^{2+\alpha}(\overline{\mathcal{X}})$, so
\begin{equation}
\label{eq:psi_m-holder}
\psi_m\in C^{2+\alpha,1+\alpha/2}
\bigl(\overline{\mathcal{X}}\times[0,\hat{T}]\bigr)
=
\overline{C}_{2+\alpha}.
\end{equation}
By Step 1, $\psi_m$ satisfies the compatibility conditions of all orders on $\partial\mathcal{X}$. By \cite[Theorem 7, Chapter 3, Section 3]{friedman2008partial}, there exists a unique classical solution
\begin{equation}
\label{eq:um-solution}
u_m\in C^{2+\alpha,1+\alpha/2}(\overline{\mathcal{X}}\times[0,\hat T])
\end{equation}
to
\[
\begin{cases}
\partial_t u_m=Lu_m, &(\bm{x},t)\in\mathcal{X}\times(0,\hat T],\\[1mm]
u_m=0, &(\bm{x},t)\in\partial\mathcal{X}\times(0,\hat T],\\[1mm]
u_m(\bm{x},0)=\phi_m(\bm{x}), &\bm{x}\in\overline{\mathcal{X}}.
\end{cases}
\]
By the weak maximum principle (\cite[Theorem 6, Chapter 2, Section 2]{friedman2008partial}),
\begin{equation}
\label{eq:um-bound}
0\le u_m\le1\qquad\text{on }\overline{\mathcal{X}}\times[0,\hat T].
\end{equation}

\noindent
\textbf{Step 3: Interior compactness.}

Let $K\Subset\mathcal{X}\times(0,\hat T]$. By the interior Schauder
estimate (\cite[Theorem 5, Chapter 3, Section 2]{friedman2008partial}),
together with~\eqref{eq:um-bound},
\[
\|u_m\|_{C^{2+\alpha,1+\alpha/2}(K)}\le C_K,
\]
where $C_K$ is independent of $m$. By the Arzel\`a--Ascoli theorem, there
exists a subsequence, still denoted by $u_m$, such that
\[
u_m\to v
\qquad\text{in }C^{2,1}_{\mathrm{loc}}
\bigl(\mathcal{X}\times(0,\hat T]\bigr)
\]
for some $v\in C^{2,1}\bigl(\mathcal{X}\times(0,\hat T]\bigr)$
 satisfying
\begin{equation}
\label{eq:v-PDE}
\partial_t v=Lv
\qquad\text{in }\mathcal{X}\times(0,\hat T].
\end{equation}

\noindent
\textbf{Step 4: Probabilistic identification.}

Fix $(\bm{x},t)\in\mathcal{X}\times(0,\hat T]$ and
$\varepsilon\in(0,t)$. For $k\geq 1$, define
\[
\tau_k
:=
\inf\Bigl\{s\geq 0\mid \operatorname{dist}(\bm{X}_{\bm{x}}(s),\partial\mathcal{X})\leq\tfrac1k\Bigr\},
\rho^{k,\varepsilon}:=\tau_k\wedge(t-\varepsilon).
\]

Before applying Itô's formula, we verify the martingale condition. Since
\[
u_m\in C^{2+\alpha,1+\alpha/2}(\overline{\mathcal{X}}\times[0,\hat T])
\]
and $\overline{\mathcal{X}}$ is compact, the function $\nabla u_m(\bm{x},r)\sigma(\bm{x})$ is continuous on the compact set $\overline{\mathcal{X}}\times[0,\hat T]$, hence uniformly bounded:
\[
M_m:=\sup_{(\bm{x},r)\in\overline{\mathcal{X}}\times[0,\hat T]}
\|\nabla u_m(\bm{x},r)\sigma(\bm{x})\|<\infty.
\]
Therefore, for any stopping time $\rho\le\hat T$,
\[
\mathbb E\left[\int_0^\rho \|\nabla u_m\sigma\|^2\,ds\right]
\le M_m^2\hat T<\infty.
\]
Thus the stochastic integral in Itô's formula is a true martingale, and taking expectations is justified without a localizing sequence.

Applying It\^o's formula to $s\mapsto u_m(\bm{X}_{\bm{x}}(s),t-s)$ up to
$\rho^{k,\varepsilon}$ gives
\[
u_m(\bm{x},t)
=
\mathbb{E}
\bigl[u_m\bigl(\bm{X}_{\bm{x}}(\rho^{k,\varepsilon}),
t-\rho^{k,\varepsilon}\bigr)\bigr],
\]
because $\partial_t u_m-Lu_m=0$. Letting $k\to\infty$ and using bounded
convergence,
\[
u_m(\bm{x},t)
=
\mathbb{E}
\bigl[\mathbf 1_{\{\tau_{\mathcal{X}}\geq t-\varepsilon\}}
u_m(\bm{X}_{\bm{x}}(t-\varepsilon),\varepsilon)\bigr].
\]
\emph{Indicator convergence at $\{\tau_{\mathcal{X}}=t\}$.} We now let
$\varepsilon\downarrow0$. On $\{\tau_{\mathcal{X}}>t\}$, we have
$\bm{X}_{\bm{x}}(t-\varepsilon)\to\bm{X}_{\bm{x}}(t)\in\mathcal{X}$ and
\[
u_m(\bm{X}_{\bm{x}}(t-\varepsilon),\varepsilon)
\to u_m(\bm{X}_{\bm{x}}(t),0)=\phi_m(\bm{X}_{\bm{x}}(t)).
\]
On $\{\tau_{\mathcal{X}}<t\}$, for $\varepsilon>0$ small enough,
$t-\varepsilon>\tau_{\mathcal{X}}$, so the indicator
$\mathbf 1_{\{\tau_{\mathcal{X}}\ge t-\varepsilon\}}$ vanishes. On
$\{\tau_{\mathcal{X}}=t\}$, we have
$\bm{X}_{\bm{x}}(t-\varepsilon)\to\bm{X}_{\bm{x}}(t)\in\partial\mathcal{X}$,
and since $\phi_m\equiv0$ on $\partial\mathcal{X}$ and $u_m$ is
continuous up to the parabolic boundary,
\[
u_m(\bm{X}_{\bm{x}}(t-\varepsilon),\varepsilon)
\to u_m(\bm{X}_{\bm{x}}(t),0)
=\phi_m(\bm{X}_{\bm{x}}(t))=0.
\]
Thus the integrand converges pointwise on $\Omega$ to
\[
\mathbf 1_{\{\tau_{\mathcal{X}}>t\}}\,\phi_m(\bm{X}_{\bm{x}}(t)).
\]
Since $0\le u_m\le1$, dominated convergence gives
\begin{equation}
\label{eq:um-representation}
u_m(\bm{x},t)
=
\mathbb E\bigl[\mathbf 1_{\{\tau_{\mathcal{X}}>t\}}\phi_m(\bm{X}_{\bm{x}}(t))\bigr].
\end{equation}
Since $\phi_m\to 1$ pointwise on $\mathcal{X}$ and $0\le\phi_m\le 1$,
another application of dominated convergence gives
\begin{equation}
\label{eq:u-identification}
v(\bm{x},t)
=
\lim_{m\to\infty}u_m(\bm{x},t)
=
\mathbb{P}(\tau_{\mathcal{X}}>t)
=
u(\bm{x},t).
\end{equation}
This limit is independent of the subsequence extracted in Step~3, so the
whole sequence converges locally in $C^{2,1}$ to $u$. In particular,
\begin{equation*}
\label{eq:u-C21}
u\in C^{2,1}\bigl(\mathcal{X}\times(0,\hat T]\bigr),
\qquad
\partial_t u=Lu
\qquad\text{in }\mathcal{X}\times(0,\hat T].
\end{equation*}

Moreover, we show that $u\in C(\mathcal X\times[0,\hat T])$.
Let $K_0\Subset\mathcal X$ be any compact set and set
$\rho:=\operatorname{dist}(K_0,\partial\mathcal X)>0$.
Then $K_0\subseteq
\{\bm{x}\in\mathcal X \mid 
\operatorname{dist}(\bm{x},\partial\mathcal X)\ge\rho\}$.
By Lemma \ref{lem:short-time-exit},
\[
\begin{aligned}
\sup_{\bm{x}\in K_0}|u(\bm{x},t)-1|
&=\sup_{\bm{x}\in K_0}
  \mathbb{P}(\tau_{\mathcal X}\le t)\\
&\le
2n\exp\left(-\frac{\rho^2}{8nS^2t}\right)
\longrightarrow0
\end{aligned}
\]
as $t\downarrow0$.
Thus $u(\cdot,t)\to1$ locally uniformly in $\mathcal X$.
Together with the established continuity for positive times,
this yields
\[
u\in C(\mathcal X\times[0,\hat T]).
\]

Further, for every $\bm{x}\in\mathcal{X}$, continuity of sample paths and
$\bm{x}\in\mathcal{X}$ open give $\tau_{\mathcal{X}}>0$ almost surely. Thus, 
\[u(\bm{x},0)=\mathbb{P}(\tau_{\mathcal{X}}>0)=1, \forall \bm{x}\in \mathcal{X}.\]

\noindent
\textbf{Step 5: Boundary continuity.}

Fix $(\bm{x}_0,t_0)\in\partial\mathcal{X}\times(0,\hat T]$. We prove that
\[
\lim_{(\bm{x},t)\to(\bm{x}_0,t_0)}u(\bm{x},t)=0.
\]
Since $u\ge 0$, it is enough to prove \[\limsup_{(\bm{x},t)\to(\bm{x}_0,t_0)}u(\bm{x},t)\le 0.\]

\noindent
\emph{\textbf{Case A}: $t_0<\hat T$.}
Choose $0<r<\min\{t_0,\hat T-t_0\}$ and set
\[
Q
:=
\bigl(B_r(\bm{x}_0)\cap\mathcal{X}\bigr)\times(t_0-r,t_0+r).
\]

By \cite{friedman2008partial}, Chapter 3, Section 4 (in particular,
the barrier construction and Theorem~8$^\prime$), there exists a local barrier
$w_Q$ such that
\[
w_Q\geq 0,
\qquad
(\partial_t-L)w_Q\geq 0
\quad\text{in }Q,
\]
with
\[
w_Q(\bm{x}_0,t_0)=0
\]
and
\[
w_Q>0
\quad\text{on }
\bigl(\overline{\mathcal{X}}\cap\overline{B_r(\bm{x}_0)}\bigr)
\times[t_0-r,t_0+r]
\setminus\{(\bm{x}_0,t_0)\}.
\]
Define the compact set
\[
\Gamma_Q
:=
\bigl(\overline{\mathcal{X}}\cap\partial B_r(\bm{x}_0)\bigr)
\times[t_0-r,t_0+r]
\;\cup\;
\bigl(\overline{\mathcal{X}}\cap\overline{B_r(\bm{x}_0)}\bigr)
\times\{t_0-r\}.
\]
Since $Q$ is small enough, $\Gamma_Q$ does not contain
$(\bm{x}_0,t_0)$, so
\[
c_Q:=\min_{\Gamma_Q}w_Q>0.
\]
For $\varepsilon>0$, choose
\[
C_\varepsilon\geq\frac{1-\varepsilon}{c_Q},
\qquad
W_\varepsilon:=\varepsilon+C_\varepsilon w_Q.
\]
By~\eqref{eq:um-bound}, $0\le u_m\le 1$. Hence
\[
u_m\leq W_\varepsilon
\quad\text{on }\Gamma_Q,
\]
and
\[
\begin{split}
u_m=0\leq W_\varepsilon
\quad&\text{on the inner lateral boundary }\\
&\bigl(\partial\mathcal{X}\cap B_r(\bm{x}_0)\bigr)
\times(t_0-r,t_0+r).
\end{split}
\]
Moreover,
\[
(\partial_t-L)(u_m-W_\varepsilon)= 
-C_\varepsilon(\partial_t-L)w_Q
\leq 0.
\]
By the weak maximum principle
(\cite[Theorem 6, Chapter 2, Section 2]{friedman2008partial}),
\begin{equation}
\label{eq:um-barrier}
u_m\leq W_\varepsilon
\qquad\text{in }Q.
\end{equation}
The barrier $W_\varepsilon$ is independent of $m$. Therefore, letting
$m\to\infty$ first in~\eqref{eq:um-barrier} gives
\[
u\leq W_\varepsilon
\qquad\text{in }Q,
\]
and only afterwards letting $(\bm{x},t)\to(\bm{x}_0,t_0)$, we obtain
\[
\limsup_{(\bm{x},t)\to(\bm{x}_0,t_0)}u(\bm{x},t)
\leq
C_\varepsilon w_Q(\bm{x}_0,t_0)+\varepsilon
=
\varepsilon.
\]
Since $\varepsilon>0$ is arbitrary,
\[
\limsup_{(\bm{x},t)\to(\bm{x}_0,t_0)}u(\bm{x},t)\leq 0.
\]
Together with $u\ge 0$, this yields
\begin{equation}
\label{eq:boundary-cont-A}
\lim_{(\bm{x},t)\to(\bm{x}_0,t_0)}u(\bm{x},t)=0.
\end{equation}

\noindent
\emph{\textbf{Case B:} $t_0=\hat T$.}
Choose $0<r<\hat T$ and set
\[
Q
:=
\bigl(B_r(\bm{x}_0)\cap\mathcal{X}\bigr)
\times(\hat T-r,\hat T].
\]
Its parabolic boundary consists of the inner lateral part
\[
\bigl(\partial\mathcal{X}\cap B_r(\bm{x}_0)\bigr)
\times(\hat T-r,\hat T),
\]
the outer lateral part
\[
\bigl(\overline{\mathcal{X}}\cap\partial B_r(\bm{x}_0)\bigr)
\times(\hat T-r,\hat T),
\]
and the bottom face
\[
\bigl(\overline{\mathcal{X}}\cap\overline B_r(\bm{x}_0)\bigr)
\times\{\hat T-r\}.
\]
The top face $t=\hat T$ is not part of the parabolic boundary.

Using the same local barrier construction as in Case~A, define
\[
\Gamma_Q
:=
\bigl(\overline{\mathcal{X}}\cap\partial B_r(\bm{x}_0)\bigr)
\times[\hat T-r,\hat T]
\;\cup\;
\bigl(\overline{\mathcal{X}}\cap\overline{B_r(\bm{x}_0)}\bigr)
\times\{\hat T-r\}.
\]
Then $\Gamma_Q$ is compact and does not contain $(\bm{x}_0,\hat T)$.
Hence
\[
c_Q:=\min_{\Gamma_Q}w_Q>0.
\]
For $\varepsilon>0$, define
\[
C_\varepsilon\ge\frac{1-\varepsilon}{c_Q},
\qquad
W_\varepsilon:=\varepsilon+C_\varepsilon w_Q.
\]

Consider the open truncated cylinder
\[
Q_{\mathrm{open}}
:=
\bigl(B_r(\bm{x}_0)\cap\mathcal{X}\bigr)
\times(\hat T-r,\hat T).
\]
On the parabolic boundary of $Q_{\mathrm{open}}$,
\[
\begin{split}
u_m\le W_\varepsilon \text{~~on }\! &\bigl(\overline{\mathcal{X}}\cap\partial B_r(\bm{x}_0)\bigr)
\times(\hat T-r,\hat T)\\
&\cup\bigl(\overline{\mathcal{X}}\cap\overline{B_r(\bm{x}_0)}\bigr)
\times\{\hat T-r\}
\end{split}
\]
and
\[
u_m=0\le W_\varepsilon
\quad\text{on }
\bigl(\partial\mathcal{X}\cap B_r(\bm{x}_0)\bigr)
\times(\hat T-r,\hat T).
\]
Moreover,
\[
(\partial_t-L)(u_m-W_\varepsilon)
=
-C_\varepsilon(\partial_t-L)w_Q
\le 0
\quad\text{in }Q_{\mathrm{open}}.
\]
The weak maximum principle (\cite[Theorem 6, Chapter 2, Section 2]{friedman2008partial}) on $Q_{\mathrm{open}}$ gives
\[
u_m\le W_\varepsilon
\qquad\text{in }Q_{\mathrm{open}}.
\]
Since both $u_m$ and $W_\varepsilon$ are continuous on the closure 
of $Q_{\mathrm{open}}$ (as $u_m \in C^{2+\alpha,1+\alpha/2}$ and 
$w_Q$ is a local barrier), we may let $t \uparrow \hat{T}$ to 
extend the inequality to the top face:
\[
u_m \le W_\varepsilon
\qquad\text{on }\bigl(B_r(\bm{x}_0)\cap\mathcal{X}\bigr)\times\{\hat{T}\}.
\]
Thus, $u_m\le W_\varepsilon$ in $Q$.

Letting $m\to\infty$, we obtain
\[
u\le W_\varepsilon
\qquad\text{in }Q.
\]
Since this bound is unconditional on $Q$, we may let
$(\bm{x},t)\to(\bm{x}_0,\hat T)$ directly:
\[
\limsup_{(\bm{x},t)\to(\bm{x}_0,\hat T)}u(\bm{x},t)
\le
C_\varepsilon w_Q(\bm{x}_0,\hat T)+\varepsilon
=
\varepsilon.
\]
Letting $\varepsilon\downarrow0$ and using $u\ge0$,
\begin{equation}
\label{eq:boundary-cont-B}
\lim_{(\bm{x},t)\to(\bm{x}_0,\hat T)}u(\bm{x},t)=0.
\end{equation}

Combining Cases A and B,
\begin{equation}
\label{eq:boundary-cont}
u\in C(\overline{\mathcal{X}}\times(0,\hat T]),
\qquad
u=0\text{ on }\partial\mathcal{X}\times(0,\hat T].
\end{equation}

Finally, for every $\bm{x}\in\mathcal{X}$, continuity of sample paths and
$\bm{x}\in\mathcal{X}$ open give $\tau_{\mathcal{X}}>0$ almost surely.
Hence
\begin{equation}
\label{eq:initial-condition}
\lim_{t\downarrow 0}u(\bm{x},t)
=
\lim_{t\downarrow 0}\mathbb{P}(\tau_{\mathcal{X}}>t)
=
1.
\end{equation}

\noindent
\textbf{Step 6: Uniqueness.}

\emph{Null set of the boundary hit event.} Step~4 gives
\[
u\in C^{2,1}\bigl(\mathcal{X}\times(0,\hat T]\bigr),
\qquad
u(\bm{x},t)=\mathbb{P}(\tau_{\mathcal{X}}>t).
\]
Hence for every fixed $\bm{x}\in\mathcal{X}$, the map
$t\mapsto u(\bm{x},t)$ is continuous on $(0,\hat T]$, and therefore so is
$t\mapsto\mathbb{P}(\tau_{\mathcal{X}}\le t)=1-u(\bm{x},t)$. Since
$s\mapsto\mathbb{P}(\tau_{\mathcal{X}}\le s)$ is nondecreasing, its
left-continuity at $t$ gives
\[
\mathbb{P}(\tau_{\mathcal{X}}<t)
=
\lim_{s\uparrow t}\mathbb{P}(\tau_{\mathcal{X}}\le s)
=
\mathbb{P}(\tau_{\mathcal{X}}\le t).
\]
Therefore
\begin{equation}
\label{eq:null-set}
\mathbb{P}(\tau_{\mathcal{X}}=t)=0
\qquad\text{for every }t\in(0,\hat T].
\end{equation}

\noindent
Let $v\in C^{2,1}\bigl(\mathcal{X}\times(0,\hat T]\bigr)
\cap C\bigl(\overline{\mathcal{X}}\times(0,\hat T]\bigr)\cap C\bigl(\mathcal{X}\times[0,\hat T]\bigr)$ be any bounded classical solution of~\eqref{eq:survival_pde}, i.e.
\[
\begin{cases}
\partial_t v=Lv
&(\bm{x},t)\in  \mathcal{X}\times(0,\hat T],\\
v=0
&(\bm{x},t)\in  \partial\mathcal{X}\times(0,\hat T],\\
\lim_{t\downarrow0}v(\bm{x},t)=1
&\bm{x}\in\mathcal{X},\\
v(\bm{x},0)=1, &\bm{x}\in \mathcal{X}.
\end{cases}
\]
Since $v$ is bounded, there exists $M>0$ such that
\[
|v(\bm{x},t)|\le M
\qquad\text{on }\mathcal{X}\times(0,\hat T].
\]

Fix $(\bm{x},t)\in\mathcal{X}\times(0,\hat T]$ and $\varepsilon\in(0,t)$.
For $k\ge1$, define
\[
\begin{split}
&\tau_k:=\inf\{s\ge0 \mid \operatorname{dist}(\bm{X}_{\bm{x}}(s),\partial\mathcal{X})\le1/k\}, \\
&\rho^{k,\varepsilon}:=\tau_k\wedge(t-\varepsilon).
\end{split}
\]

\noindent
\emph{Martingale condition.} Set
\[
\overline{\mathcal{X}}_k:=\{\bm{y}\in\mathcal{X} \mid 
\operatorname{dist}(\bm{y},\partial\mathcal{X})\ge1/k\}.
\]
The set $\overline{\mathcal{X}}_k\times[\varepsilon,t]$ is a compact
subset of $\mathcal{X}\times(0,\hat T]$, so the map
$(\bm{y},r)\mapsto\nabla v(\bm{y},r)\sigma(\bm{y})$ is continuous there,
hence uniformly bounded:
\[
M_{k,\varepsilon}
:=
\sup_{(\bm{y},r)\in\overline{\mathcal{X}}_k\times[\varepsilon,t]}
\|\nabla v(\bm{y},r)\sigma(\bm{y})\|<\infty.
\]
Therefore $\mathbb E\!\left[ \int_0^{\rho^{k,\varepsilon}} \|\nabla v\,\sigma\|^2\,ds \right] \le M_{k,\varepsilon}^2\,t<\infty$,
so the stochastic integral in It\^o's formula is a true martingale.

\noindent
\emph{It\^o's formula.}
Applying It\^o's formula to
$s\mapsto v(\bm{X}_{\bm{x}}(s),t-s)$ on
$[0,\rho^{k,\varepsilon}]$ and using $\partial_t v=Lv$,
\begin{equation}
\label{eq:feynman-kac-v}
v(\bm{x},t)
=
\mathbb E\bigl[
v\bigl(\bm{X}_{\bm{x}}(\rho^{k,\varepsilon}),
t-\rho^{k,\varepsilon}\bigr)
\bigr].
\end{equation}

\noindent \emph{Passage to the limit.}
Let $(k_n,\varepsilon_n)$ be any sequence with $k_n\to\infty$ and
$\varepsilon_n\downarrow0$, and set $\rho_n:=\rho^{k_n,\varepsilon_n}$.
Then $\rho_n\to\tau_{\mathcal{X}}\wedge t$ almost surely. By~\eqref{eq:null-set},
the event $\{\tau_{\mathcal{X}}=t\}$ is negligible, so it suffices to
verify convergence on the two complementary events.

\begin{enumerate}
\item[(i)] On $\{\tau_{\mathcal{X}}>t\}$, we have $\rho_n=t-\varepsilon_n$
for $n$ large. By path continuity,
$\bm{X}_{\bm{x}}(t-\varepsilon_n)\to\bm{X}_{\bm{x}}(t)\in\mathcal{X}$.
The path image $K_\omega := \{\bm{X}_{\bm{x}}(s)\mid 0\le s\le t\}$ is a 
compact subset of $\mathcal{X}$. Because $v$ is continuous on the compact 
set $K_\omega \times [0, t]$, it is uniformly continuous there. Therefore, 
since $\bm{X}_{\bm{x}}(t-\varepsilon_n)\to\bm{X}_{\bm{x}}(t)$ and 
$\varepsilon_n \to 0$,
\[
v\bigl(\bm{X}_{\bm{x}}(t-\varepsilon_n),\varepsilon_n\bigr)
\longrightarrow
v\bigl(\bm{X}_{\bm{x}}(t),0\bigr) = 1.
\]

\item[(ii)] On $\{\tau_{\mathcal{X}}<t\}$, we have $\rho_n=\tau_{k_n}$
for $n$ large, with $\tau_{k_n}\uparrow\tau_{\mathcal{X}}$. By path
continuity,
$\bm{X}_{\bm{x}}(\tau_{k_n})\to\bm{X}_{\bm{x}}(\tau_{\mathcal{X}})\in\partial\mathcal{X}$,
and since $t-\tau_{\mathcal{X}}>0$ and $v=0$ on
$\partial\mathcal{X}\times(0,\hat T]$,
\[
v\bigl(\bm{X}_{\bm{x}}(\tau_{k_n}),t-\tau_{k_n}\bigr)
\longrightarrow
v\bigl(\bm{X}_{\bm{x}}(\tau_{\mathcal{X}}),t-\tau_{\mathcal{X}}\bigr)=0.
\]
\end{enumerate}
Combining the two events,
\[
v\bigl(\bm{X}_{\bm{x}}(\rho_n),t-\rho_n\bigr)
\longrightarrow
\mathbf 1_{\{\tau_{\mathcal{X}}>t\}}
\qquad\text{almost surely}.
\]
Since $|v|\le M$, the dominated convergence theorem applied
to~\eqref{eq:feynman-kac-v} yields
\begin{equation}
\label{eq:v-representation}
v(\bm{x},t)
=
\mathbb{P}(\tau_{\mathcal{X}}>t)
=
u(\bm{x},t).
\end{equation}

\noindent
\emph{Uniqueness.}
Equation~\eqref{eq:v-representation} holds for every bounded classical
solution $v$ and every $(\bm{x},t)\in\mathcal{X}\times(0,\hat T]$. Hence
every bounded classical solution coincides with the function $u$
constructed in Step~4:
\[
v(\bm{x},t)=u(\bm{x},t)=\mathbb{P}(\tau_{\mathcal{X}}>t).
\]
In particular, if $v_1$ and $v_2$ are two bounded classical solutions,
then $v_1=v_2$. This proves uniqueness.

\noindent
\textbf{Part ii: Positive-time regularity up to the boundary.}

Fix $\delta\in(0,\hat T)$. We show
\begin{equation}
\label{eq:um-holder-bound}
\|u_m\|_{C^{2+\alpha,1+\alpha/2}
(\overline{\mathcal{X}}\times[\delta,\hat T])}\le C_\delta,
\end{equation}
with $C_\delta$ independent of $m$.

\noindent
\emph{Interior estimate.}
Let $K\Subset\mathcal{X}\times[\delta,\hat T]$. By~\eqref{eq:um-bound}
and the interior Schauder estimate,
\begin{equation}
\label{eq:interior-holder}
\|u_m\|_{C^{2+\alpha,1+\alpha/2}(K)}\le C_K,
\end{equation}
with $C_K$ independent of $m$.

\noindent
\emph{Boundary estimate.}
Let $(\bm{x}_0,t_0)\in\partial\mathcal{X}\times[\delta,\hat{T}]$. Choose
$r>0$ sufficiently small so that the parabolic neighbourhood
\[
Q :=
\begin{cases}
\bigl(B_r(\bm{x}_0)\cap\mathcal{X}\bigr)\times(t_0-r,t_0+r),
& t_0<\hat{T},\\[1mm]
\bigl(B_r(\bm{x}_0)\cap\mathcal{X}\bigr)\times(\hat{T}-r,\hat{T}],
& t_0=\hat{T},
\end{cases}
\]
satisfies
\[
\overline Q\subset\overline{\mathcal{X}}\times[\delta/2,\hat{T}].
\]
Set
\[
J :=
\begin{cases}
(t_0-r,\,t_0+r), & t_0<\hat{T},\\[1mm]
(\hat{T}-r,\,\hat{T}], & t_0=\hat{T}.
\end{cases}
\]
For every $m$, the function $u_m$ belongs to
$C^{2+\alpha,1+\alpha/2}(\overline{\mathcal{X}}\times[0,\hat{T}])$
by~\eqref{eq:um-solution}. Hence the boundary Schauder estimate
(\cite[Theorem~4, Chapter~4, Section~7]{friedman2008partial}) applies to
$u_m$ in a smaller parabolic neighbourhood $Q'\Subset Q$ of
$(\bm{x}_0,t_0)$ that reaches the boundary.

Since
\[
\partial_t u_m - L u_m = 0
\quad\text{in } Q,
\quad
u_m = 0
\quad\text{on }
\bigl(\partial\mathcal{X}\cap B_r(\bm{x}_0)\bigr)\times J,
\]
the only quantity entering the boundary Schauder estimate that may depend on
$m$ is the $L^\infty$-norm of $u_m$ itself. By~\eqref{eq:um-bound},
\[
\|u_m\|_{L^\infty(Q)} \le 1
\qquad\text{for all }m.
\]
Applying the boundary Schauder estimate on the smaller neighbourhood
$Q'\Subset Q$ therefore yields
\begin{equation}
\label{eq:boundary-holder}
\|u_m\|_{C^{2+\alpha,1+\alpha/2}(Q')}\le C_Q,
\end{equation}
where the constant $C_Q$ depends only on $Q$, the coefficients, and the
uniform $L^\infty$-bound $1$, but not on $m$.

\noindent
\textbf{Global bound.}
Since $\overline{\mathcal{X}}\times[\delta,\hat T]$ is compact, finitely
many interior and boundary neighbourhoods cover this set. Combining the
corresponding local estimates~\eqref{eq:interior-holder}
and~\eqref{eq:boundary-holder}, and using the standard localization
property of parabolic H\"older norms, a finite
collection of local $C^{2+\alpha,1+\alpha/2}$-bounds, uniform in $m$,
yields a global $C^{2+\alpha,1+\alpha/2}$-bound on the compact set
$\overline{\mathcal{X}}\times[\delta,\hat{T}]$:
\begin{equation}
\label{eq:global-holder}
\|u_m\|_{C^{2+\alpha,1+\alpha/2}
(\overline{\mathcal{X}}\times[\delta,\hat T])}\le C_\delta,
\end{equation}
where $C_\delta$ is independent of $m$.

\noindent
\textbf{Passing to the limit.}
By the Arzel\`a--Ascoli theorem applied to $u_m$ and its derivatives on the compact set $\overline{\mathcal{X}}\times[\delta,\hat T]$, there is a subsequence converging in
$C^{2,1}(\overline{\mathcal{X}}\times[\delta,\hat T])$. Since the pointwise limit is the already identified function $u$ (by Step~4), the whole sequence converges. Passing to the limit in~\eqref{eq:global-holder} and using the lower semicontinuity of the H\"older seminorms under pointwise convergence, we obtain
$\|u\|_{C^{2+\alpha,1+\alpha/2}
(\overline{\mathcal{X}}\times[\delta,\hat T])}\le C_\delta$. Hence, $u\in C^{2+\alpha,1+\alpha/2}
\bigl(\overline{\mathcal{X}}\times[\delta,\hat T]\bigr)$.

This completes the proof.  \qed
\end{pf}

\begin{lemma}[Polynomial approximation in $C^{2,1}$]
\label{lem:poly-approx}
Let $U\subset\mathbb R^{1+n}$ be open and let $K\subset U$ be compact. Write a point of $\mathbb R^{1+n}$ as $(\bm{x},t)$. Suppose $f\in C^{2,1}(U)$. Then there exists a sequence of polynomials $p_k\in\mathbb R[\bm{x},t]$ such that
\begin{equation}
\label{eq:c12-poly-approx}
\|p_k-f\|_{C^{2,1}(K)}
\longrightarrow 0,
\end{equation}
where
\begin{equation}
\label{eq:c12-norm}
\|g\|_{C^{2,1}(K)}:=
\max\left\{
\begin{split}
&\sup_{(\bm{x},t)\in K}|g(\bm{x},t)|,\\
&\sup_{(\bm{x},t)\in K}|\partial_t g(\bm{x},t)|,\\
&\sup_{(\bm{x},t)\in K}\|\nabla_{\bm{x}}g(\bm{x},t)\|,\\
&\sup_{(\bm{x},t)\in K}\|\bm H_{\bm{x}}g(\bm{x},t)\|_F
\end{split}
\right\}.
\end{equation}
\end{lemma}
\begin{pf}
Since $K\subset U$ is compact and $U$ is open, there exists a bounded
open set $U_1\subset\mathbb R^{1+n}$ such that
\[
K\subset U_1,
\qquad
\overline{U_1}\subset U.
\]
Set
\[
d:=\operatorname{dist}(\overline{U_1},U^c)>0.
\]
In particular, there exist $\delta_1,\delta_2>0$ such that the
$\delta_1$-neighborhood of $K$ is contained in $U_1$, and the
$\delta_2$-neighborhood of $\overline{U_1}$ is contained in $U$.

\noindent\textbf{Step 1: Mollification.}
Let $\rho\in C_c^\infty(\mathbb R^{1+n})$ be a standard mollifier
satisfying $\operatorname{supp}\rho\subset B_1(\bm{0})$ and
$\int_{\mathbb R^{1+n}}\rho(\bm z)\,d\bm z=1$. For $r>0$, set
\[
\rho_r(\bm z)=r^{-(1+n)}\rho(\bm z/r).
\]
For $0<r<d$, define the local convolution on $U_1$ by
\begin{equation}
\label{eq:local-convolution}
f_r(\bm z)
=
\int_{B_r(0)}
\rho_r(\bm y)\,f(\bm z-\bm y)\,d\bm y,
\qquad \bm z\in U_1.
\end{equation}
Since $r<d=\operatorname{dist}(\overline{U_1},U^c)$, every value of $f$
appearing in~\eqref{eq:local-convolution} is well-defined. Consequently,
$f_r\in C^\infty(U_1)$.

For $r$ sufficiently small, the $r$-neighborhood of $K$ is contained
in $U_1$. Since $f$, $\partial_t f$, $\nabla_{\bm{x}}f$, and
$\bm H_{\bm{x}}f$ are continuous on the compact set
$\overline{U_1}$, they are uniformly continuous there. The standard
convergence properties of mollification therefore give
\[
\|f_r-f\|_{C^{2,1}(K)}\longrightarrow 0
\qquad\text{as }r\downarrow0.
\]
Choose a sequence $r_j\downarrow0$ such that $\|f_{r_j}-f\|_{C^{2,1}(K)}<\frac{1}{2j}$, and set $f_j:=f_{r_j}$.

\noindent\textbf{Step 2: Polynomial approximation.}
Consider the algebra
\[
\mathcal P
:=
\{q|_{U_1} \mid q\in\mathbb R[t,\bm{x}]\}
\subset C^\infty(U_1).
\]
This algebra contains the constants and separates distinct points of $U_1$. Moreover, the coordinate functions $t,x_1,\ldots,x_n$ belong to $\mathcal P$, and their differentials span the cotangent space at every point of $U_1$. Thus $\mathcal P$ separates tangent vectors.

By Nachbin's smooth Stone--Weierstrass theorem
\cite{nachbin1949algebres}, $\mathcal P$ is dense in $C^\infty(U_1)$ in the compact-open $C^\infty$ topology. This topology controls the uniform convergence of all derivatives on every compact subset of $U_1$, and in particular implies approximation in the $C^{2,1}(K)$
norm.

Consequently, for each $j$, there exists a polynomial
$p_j\in\mathbb R[\bm{x},t]$ such that $\|p_j-f_j\|_{C^{2,1}(K)}<\frac{1}{2j}$.

\noindent\textbf{Step 3: Diagonal argument.}
By the triangle inequality,
\[
\begin{aligned}
\|p_j-f\|_{C^{2,1}(K)}
&\le
\|p_j-f_j\|_{C^{2,1}(K)}
+
\|f_j-f\|_{C^{2,1}(K)}\\
&<\frac{1}{2j}+\frac{1}{2j}
=
\frac{1}{j}.
\end{aligned}
\]
Hence $\|p_j-f\|_{C^{2,1}(K)}\longrightarrow0
\qquad\text{as }j\to\infty$,
which proves~\eqref{eq:c12-poly-approx}. \qed
\end{pf}

\section*{Appendix B}

\begin{proposition}
The unique classical solution to the parabolic initial-boundary value problem
\begin{equation}
\label{eq:benchmark-pde}
\begin{cases}
\partial_t u=\dfrac{\sigma^2}{2}\,\partial_{xx}u,
&(x,t)\in(-1,1)\times(0,T],\\[1mm]
u(\pm1,t)=0,
&t\in(0,T],\\[1mm]
\displaystyle\lim_{t\downarrow0}u(x,t)=1,
&x\in(-1,1).
\end{cases}
\end{equation}
admits the rapidly convergent eigenfunction
expansion at time $t=T$
\begin{equation*}
\label{eq:benchmark-survival}
\begin{split}
&u(x_0,T)=\\
&\frac{4}{\pi}
\sum_{k=0}^{\infty}
\frac{(-1)^k}{2k+1}
\cos\!\left(\frac{(2k+1)\pi x_0}{2}\right)
\exp\!\left(-\frac{(2k+1)^2\pi^2\sigma^2T}{8}\right).
\end{split}
\end{equation*}
Consequently, the finite-horizon exit probability is
\begin{equation*}
\label{eq:benchmark-exit}
\begin{split}
&\mathbb{P}_{\mathrm{exit}}(\tau_{\mathcal{X}}(x_0)\le T)\\=&
1-u(x_0,T)\\
=&1-\frac{4}{\pi}
\sum_{k=0}^{\infty}
\frac{(-1)^k}{2k+1}a(k,x_0)b(k,x_0,T).
\end{split}
\end{equation*}
where $a(k,x_0)=\cos\!\left(\frac{(2k+1)\pi x_0}{2}\right)$ and 
$b(k,x_0,T)=\exp\!\left(-\frac{(2k+1)^2\pi^2\sigma^2T}{8}\right)$. 
\end{proposition}
\begin{pf}
We solve the parabolic initial-boundary value problem
\begin{equation}
\label{eq:ibvp}
\begin{cases}
\partial_t u=\dfrac{\sigma^2}{2}\,\partial_{xx}u,
&(x,t)\in(-1,1)\times(0,T],\\[1mm]
u(\pm1,t)=0,
&t\in(0,T],\\[1mm]
\displaystyle\lim_{t\downarrow0}u(x,t)=1,
&x\in(-1,1).
\end{cases}
\end{equation}

\smallskip
\noindent\textbf{Step 1: Separation of variables.}
Seek solutions of the form $u(x,t)=X(x)\,\Theta(t)$. Substituting
into~\eqref{eq:ibvp} and dividing by $X(x)\Theta(t)$ gives
\[
\frac{\Theta'(t)}{\Theta(t)}
=
\frac{\sigma^2}{2}\frac{X''(x)}{X(x)}
=
-\lambda,
\]
where $\lambda$ is the separation constant. This yields two ordinary
differential equations,
\[
\Theta'(t)+\lambda\,\Theta(t)=0,
\qquad
X''(x)+\mu X(x)=0,
\qquad
\mu:=\frac{2\lambda}{\sigma^2}.
\]

\smallskip
\noindent\textbf{Step 2: Spatial eigenvalue problem.}
The boundary conditions $u(\pm1,t)=0$ imply $X(\pm1)=0$. We solve
\[
X''(x)+\mu X(x)=0,
\qquad
X(\pm1)=0.
\]

\emph{Case $\mu\le 0$.} The general solution is a combination of
exponentials or a linear function; the boundary conditions force
$X\equiv 0$.

\emph{Case $\mu>0$.} Write $\mu=k^2$ with $k>0$. Then
\[
X(x)=A\cos(kx)+B\sin(kx),
\]
and the boundary conditions give
\[
A\cos k+B\sin k=0,
\qquad
A\cos k-B\sin k=0.
\]
Adding and subtracting,
\[
2A\cos k=0,
\qquad
2B\sin k=0.
\]
Hence either $B=0$ and $\cos k=0$, or $A=0$ and $\sin k=0$.

\begin{itemize}
\item If $B=0$ and $\cos k=0$, then $k=\dfrac{(2m+1)\pi}{2}$,
$m\ge0$, giving
\[
\varphi_m(x)=\cos\!\left(\frac{(2m+1)\pi x}{2}\right).
\]
\item If $A=0$ and $\sin k=0$, then $k=m\pi$, $m\ge1$, giving
\[
\psi_m(x)=\sin(m\pi x).
\]
\end{itemize}
Both families satisfy the Dirichlet boundary conditions and, together, form
a complete orthogonal basis of $L^2(-1,1)$.

\smallskip
\noindent\textbf{Step 3: Time equation.}
For each eigenvalue $\mu$, the time equation reads
\[
\Theta'(t)+\lambda\,\Theta(t)=0,
\qquad
\lambda=\frac{\sigma^2\mu}{2},
\]
so $\Theta(t)=e^{-\lambda t}$.

\smallskip
\noindent\textbf{Step 4: General solution.}
By linearity and completeness, the general solution of~\eqref{eq:ibvp} is
\[
u(x,t)=\sum_{k=0}^{\infty} a_k\,\varphi_k(x)\,e^{-\lambda_k t}
+\sum_{m=1}^{\infty} b_m\,\psi_m(x)\,e^{-\tilde\lambda_m t},
\]
where
\[
\lambda_k=\frac{(2k+1)^2\pi^2\sigma^2}{8},
\qquad
\tilde\lambda_m=\frac{m^2\pi^2\sigma^2}{2}.
\]

\smallskip
\noindent\textbf{Step 5: Imposing the initial condition.}
The initial datum $u(x,0)=1$ on $(-1,1)$ requires
\[
1=\sum_{k=0}^{\infty} a_k\,\varphi_k(x)
+\sum_{m=1}^{\infty} b_m\,\psi_m(x).
\]
Using the orthogonality relations
\[
\begin{split}
&\int_{-1}^{1}\varphi_k(x)\varphi_\ell(x)\,dx
=
\begin{cases}
1, & k=\ell,\\
0, & k\neq \ell,
\end{cases}\\
&\int_{-1}^{1}\psi_m(x)\psi_n(x)\,dx
=
\begin{cases}
1, & m=n,\\
0, & m\neq n,
\end{cases}
\end{split}
\]
the coefficients are computed as follows.

\emph{Cosine coefficients.}
For $\varphi_k(x)=\cos\!\left(\dfrac{(2k+1)\pi x}{2}\right)$,
\[
\begin{split}
a_k
&=
\int_{-1}^{1}\cos\!\left(\frac{(2k+1)\pi x}{2}\right)dx\\
&=
\frac{2}{(2k+1)\pi}\sin\!\left(\frac{(2k+1)\pi x}{2}\right)\Bigg|_{-1}^{1}.
\end{split}
\]
Since
\[
\sin\!\left(\frac{(2k+1)\pi}{2}\right)=(-1)^k, 
\sin\!\left(-\frac{(2k+1)\pi}{2}\right)=-(-1)^k,
\]
we obtain
\[
a_k
=
\frac{4(-1)^k}{(2k+1)\pi}.
\]

\emph{Sine coefficients.}
For $\psi_m(x)=\sin(m\pi x)$,
\[
\begin{split}
b_m
&=
\int_{-1}^{1}\sin(m\pi x)\,dx=
-\frac{1}{m\pi}\cos(m\pi x)\Bigg|_{-1}^{1}\\
&=
-\frac{1}{m\pi}\bigl[(-1)^m-(-1)^m\bigr]=
0.
\end{split}
\]
So the sine modes do not contribute.

\smallskip
\noindent\textbf{Step 6: Final formula.}
Combining the spatial eigenfunctions, the time factors, and the Fourier
coefficients yields the rapidly convergent eigenfunction expansion
\begin{equation}
\label{eq:survival-expansion}
\begin{split}
u(x_0,T)=\frac{4}{\pi}
\sum_{k=0}^{\infty}
\frac{(-1)^k}{2k+1}a(k,x_0) b(k,x_0,T)
\end{split}
\end{equation}
where $a(k,x_0)=\cos\!\left(\frac{(2k+1)\pi x_0}{2}\right)$ and $b(k,x_0,T)=
\exp\!\left(-\frac{(2k+1)^2\pi^2\sigma^2T}{8}\right)$.
Consequently, the finite-horizon exit probability is
\begin{equation}
\label{eq:exit-expansion}
\begin{split}
\mathbb{P}(\tau_{\mathcal X}\le T)=
1-u(x_0,T)
\end{split}
\end{equation}

\smallskip
\noindent\textbf{Step 7: Justification of convergence and uniqueness.}
\begin{itemize}
\item \emph{Convergence.} For every $\delta>0$, the terms
of~\eqref{eq:survival-expansion} decay like
\[
\exp\!\left(-\frac{(2k+1)^2\pi^2\sigma^2\delta}{8}\right),
\]
which is summable. Hence the series converges absolutely and uniformly on
$[-1,1]\times[\delta,T]$, and it can be differentiated term by term in
$x$ and $t$. Each term solves the heat equation, so the sum is a
classical solution on $(-1,1)\times(0,T]$.

\item \emph{Initial condition.} As $t\downarrow0$, the series
converges to the Fourier series of the constant function $1$, which
equals $1$ on $(-1,1)$ by the classical theory of Fourier series.

\item \emph{Uniqueness.} If $u_1$ and $u_2$ are two classical
solutions with the same initial and boundary data, then $w=u_1-u_2$
solves the heat equation with $w=0$ on the parabolic boundary. The
parabolic maximum principle forces $w\equiv0$.
\end{itemize}
Therefore \eqref{eq:ibvp} has a unique classical solution, given
by~\eqref{eq:survival-expansion}. \qed
\end{pf}

\begin{proposition} 
For Example \ref{ex2},
\begin{equation*}
\label{eq:benchmark-coupled-exit}
\mathbb{P}_{\mathrm{exit}}(x_0)
=
1-
\sum_{n=1}^\infty
c_n\,
{}_1F_1\!\left(
-\alpha_n;1;
\frac{\theta\|x_0\|^2}{\sigma^2}
\right)
e^{-2\theta\alpha_nT},
\end{equation*}
with coefficients
\[
c_n
=
\frac{
\int_0^R
\phi_n(s)\,
s\,e^{-\theta s^2/\sigma^2}\,ds
}{
\int_0^R
\phi_n(s)^2\,
s\,e^{-\theta s^2/\sigma^2}\,ds
}.
\]    
\end{proposition}
\begin{pf}
\smallskip
\noindent\textbf{Step 1: Radial reduction.}
Let $v(\bm{x})=\varphi(r)$ with $r=\|\bm{x}\|$. In polar coordinates,
\[
\nabla v=\varphi'(r)\frac{\bm{x}}{r},
\qquad
\Delta v=\varphi''(r)+\frac{1}{r}\varphi'(r).
\]
The drift is $\bm{b}(\bm{x})=(-\theta x_1-\omega x_2,\,
\omega x_1-\theta x_2)$. Hence
\[
\begin{split}
\bm{b}\cdot\nabla v&=
(-\theta x_1-\omega x_2)\frac{\varphi'(r)}{r}x_1
+
(\omega x_1-\theta x_2)\frac{\varphi'(r)}{r}x_2\\
&=-\theta r\varphi'(r),
\end{split}
\]
because the $\omega$-terms cancel. Therefore the generator reduces to
\[
L v
=
\frac{\sigma^2}{2}\Bigl(\varphi''(r)+\frac{1}{r}\varphi'(r)\Bigr)
-\theta r\varphi'(r).
\]

\smallskip
\noindent\textbf{Step 2: Separation of variables.}
The safety probability $u(r,t)=\mathbb{P}(\tau_{\mathcal{X}}>t)$
satisfies the backward equation $\partial_t u=Lu$ with the given boundary
and initial conditions. Assume $u(r,t)=R(r)T(t)$. Then
\[
\frac{T'(t)}{T(t)}
=
\frac{\frac{\sigma^2}{2}\bigl(R''(r)+\frac{1}{r}R'(r)\bigr)
-\theta r R'(r)}{R(r)}
=
-\lambda,
\]
so $T(t)=e^{-\lambda t}$ and the spatial equation becomes
\[
\frac{\sigma^2}{2}\Bigl(R''+\frac{1}{r}R'\Bigr)
-\theta r R'
=
-\lambda R.
\]

\smallskip
\noindent\textbf{Step 3: Spatial eigenfunctions.}
Set $z=\theta r^2/\sigma^2$ and write $R(r)=f(z)$. A direct computation
gives
\[
\frac{dz}{dr}=\frac{2\theta r}{\sigma^2},
\qquad
\frac{d^2z}{dr^2}=\frac{2\theta}{\sigma^2}.
\]
Hence
\[
R'(r)=f'(z)\frac{2\theta r}{\sigma^2},
\qquad
R''(r)=f''(z)\frac{4\theta^2 r^2}{\sigma^4}
+f'(z)\frac{2\theta}{\sigma^2}.
\]
Substituting into the spatial equation and simplifying yields Kummer's
equation
\[
z f''(z)+(1-z)f'(z)+\frac{\lambda}{2\theta}f(z)=0.
\]
Writing $\lambda_n=2\theta\alpha_n$, the regular solution at $z=0$ is
\[
\phi_n(r)={}_1F_1(-\alpha_n;1;z),
\]
where
\[
{}_1F_1(a;b;z)
=
\sum_{k=0}^{\infty}\frac{(a)_k}{(b)_k}\frac{z^k}{k!},
\qquad
(a)_k=a(a+1)\cdots(a+k-1).
\]
The Dirichlet boundary condition $u(R,t)=0$ gives
\[
{}_1F_1\!\left(-\alpha_n;1;\frac{\theta R^2}{\sigma^2}\right)=0,
\]
which determines the positive roots $\alpha_n$. The condition
$u_r(0,t)=0$ is automatically satisfied because $\phi_n'(0)=0$.
The corresponding eigenvalues are $\lambda_n=2\theta\alpha_n$.

\smallskip
\noindent\textbf{Step 4: Orthogonality and coefficients.}
The spatial operator in Sturm--Liouville form is
\[
\Bigl(r e^{-\theta r^2/\sigma^2}R'\Bigr)'
+
\frac{2\lambda}{\sigma^2}
r e^{-\theta r^2/\sigma^2}R=0.
\]
Therefore the eigenfunctions are orthogonal with respect to the weight
$w(r)=r e^{-\theta r^2/\sigma^2}$:
\[
\int_0^R\phi_n(r)\phi_m(r)\,w(r)\,dr=0,
\qquad n\neq m.
\]
Expanding the initial datum $u(r,0)=1$ in this basis gives
\[
u(r,t)=\sum_{n=1}^{\infty}c_n\phi_n(r)e^{-2\theta\alpha_n t},
\]
with
\[
c_n
=
\frac{\displaystyle\int_0^R\phi_n(s)\,s\,e^{-\theta s^2/\sigma^2}\,ds}
{\displaystyle\int_0^R\phi_n(s)^2\,s\,e^{-\theta s^2/\sigma^2}\,ds}.
\]

\smallskip
\noindent\textbf{Step 5: Final formula.}
Substituting $z=\theta r^2/\sigma^2$ and evaluating at $r=\|\bm{x}_0\|$,
$t=T$, we obtain
\[
\begin{split}
&\mathbb{P}_{\mathrm{exit}}(\bm{x}_0)\\
=&1-u(\|\bm{x}_0\|,T)\\
=&1-
\sum_{n=1}^{\infty}
c_n\,
{}_1F_1\!\left(
-\alpha_n;1;
\frac{\theta\|\bm{x}_0\|^2}{\sigma^2}
\right)
e^{-2\theta\alpha_n T}.
\end{split}
\]
This is the exact spectral representation of the exit probability. \qed
\end{pf}

\end{document}